\documentclass[prb,twocolumn,showpacs,preprintnumbers,amsmath,amssymb,citeautoscript]{revtex4-1}

\usepackage{graphicx}
\usepackage{amssymb, amsfonts, amsmath}
\usepackage{dcolumn}
\usepackage{bm}
\usepackage{color}

\newlength{\figwidth}
\begin{document}


\title{From Kondo Lattices to Kondo Superlattices}

\author{Masaaki~Shimozawa$^{1}$}
\author{Swee~K.~Goh$^{2}$}
\author{Takasada~Shibauchi$^{3}$}
\author{Yuji~Matsuda$^{4}$}

\affiliation{
$^1$The Institute for Solid State Physics (ISSP), The University of Tokyo, Kashiwa, Chiba 277-8581, Japan\\$^2$Department of Physics, The Chinese University of Hong Kong, Shatin, New Territories, Hong Kong, China\\
$^3$Department of Advanced Materials Science, University of Tokyo, Kashiwa, Chiba 277-8561, Japan\\
$^4$Department of Physics, Kyoto University, Kyoto 606-8502, Japan}

\date{\today}

\begin{abstract}
Realization of new classes of ground states in strongly correlated electron systems continues to be a major issue in condensed matter physics. Heavy fermion materials, whose electronic structure is essentially three-dimensional, are one of the most suitable systems for obtaining novel electronic states because of their intriguing properties associated with many-body effects. Recently, a state-of-the-art molecular beam epitaxy technique was developed to reduce the dimensionality of heavy electron systems by fabricating artificial superlattices that include heavy fermion compounds; this approach can produce a new type of electronic state in two-dimensional (2D) heavy fermion systems. In artificial superlattices of the antiferromagnetic heavy fermion compound CeIn$_3$ and the conventional metal LaIn$_3$, the magnetic order is suppressed by a reduction in the thickness of the CeIn$_3$ layers. In addition, the 2D confinement of heavy fermions leads to enhancement of the effective electron mass and deviation from the standard Fermi liquid electronic properties, which are both associated with the dimensional tuning of quantum criticality. In the superconducting superlattices of the heavy fermion superconductor CeCoIn$_5$ and nonmagnetic metal YbCoIn$_5$, signatures of superconductivity are observed even at the thickness of one unit-cell layer of CeCoIn$_5$. The most remarkable feature of this 2D heavy fermion superconductor is that the thickness reduction of the CeCoIn$_5$ layers changes the temperature and angular dependencies of the upper critical field significantly. This result is attributed to a substantial suppression of the Pauli pair-breaking effect through the local inversion symmetry breaking at the interfaces of CeCoIn$_5$ block layers. The importance of the inversion symmetry breaking in this system has also been supported by site-selective nuclear magnetic resonance spectroscopy, which can resolve spectroscopic information from each layer separately, even within the same CeCoIn$_5$ block layer. In addition, recent experiments involving CeCoIn$_5$/YbCoIn$_5$ superlattices have shown that the degree of the inversion symmetry breaking and, in turn, the Rashba splitting are controllable, offering the prospect of achieving even more fascinating superconducting states. Thus, these Kondo superlattices pave the way for the exploration of unconventional metallic and superconducting states.
\end{abstract}


\maketitle
\tableofcontents

\section{Introduction}
Strongly interacting quantum many-body systems are an important issue in modern physics. In condensed matter systems, such complex interactions between particles are responsible for intricate interplay between charge, spin, and orbital degrees of freedom, giving rise to a variety of intriguing ground states such as unconventional superconductivity, magnetically ordered states, and states exhibiting interesting charge and orbital orderings. These phases often coexist and compete with one another, resulting in complicated phase diagrams. Transitions between these phases can generally be driven by external parameters, including pressure, chemical doping, and magnetic fields. The case of driving a second-order phase transition at zero temperature is of particular interest\cite{Sachdev2000,Gegenwart2008,Shibauchi2014}. In the vicinity of such a zero-temperature transition point, called the quantum critical point (QCP), quantum fluctuations induce substantial modifications of the quasiparticle masses and scattering cross sections of the Fermi liquid. These modifications cause striking deviations from the conventional Landau Fermi liquid description of metallic systems. Furthermore, superconductivity is often observed in the phase space near the QCP. Thus, strongly correlated electron systems, which exhibit many fascinating physical phenomena and offer suitable platforms for exploring exotic physics resulting from quantum criticality, have attracted considerable attention from condensed matter researchers.

Among the existing strongly correlated materials, the ground state with the strongest electron correlation effects is achieved in heavy fermion compounds containing 4$f$ or 5$f$ electrons \cite{Stewart1984}. In these systems, the $f$ electrons that can couple with conduction electrons form a dense periodic array. This array is called a Kondo lattice, and it is usually three dimensional (3D). These $f$ electrons, which are localized at high temperatures, hybridize with the conduction electrons at low temperatures, thereby forming a very narrow conduction band. The quasiparticle effective mass of the narrow band is substantially enhanced, sometimes reaching a value as large as 1000 times the free-electron mass. The term ``heavy fermion" is derived from this fact. In such heavy fermion systems, the magnetic moment of localized $f$ electrons is quenched through the Kondo effect, leading to a nonmagnetic state. By contrast, the localized $f$ electron moments interact with each other via the Ruderman--Kittel--Kasuya--Yoshida (RKKY) interaction, which promotes magnetic ordering. The competition between the Kondo effect and the RKKY interaction is a key element in determining whether the ground state is a non-magnetic metal or a magnetically ordered state. Moreover, the interplay between spin, valence, and orbital degrees of freedom of the $f$ electrons affects the electronic and magnetic states, leading to a rich variety of phenomena such as the coexistence of superconductivity and ferromagnetism \cite{Aoki2001,Huy2007}, multiple superconducting phases with different order parameter symmetries \cite{Joynt2002}, reentrant superconductivity \cite{Levy2005}, multipolar ordering \cite{Matsubayashi2012,Tsujimoto2014}, and mixed valence states \cite{Yuan2003}. Thus, the 3D Kondo lattice plays an important role in heavy fermion physics.

Generally, in low spatial dimensions, many-body correlation effects originating from the Coulomb interaction between electrons become more important and complex. Moreover, both thermal and quantum fluctuations are largely enhanced with a reduction in dimensionality, thereby expanding the critical regions near the QCP. Thus, many-body effects that do not exist in three dimensions are expected to prevail in low-dimensional systems. Indeed, two-dimensional (2D) systems exhibit anomalous properties such as high-$T_c$ superconductivity in cuprates \cite{Bednorz1986,Damascelli2003} and iron pnictides \cite{Kamihara2006,Stewart2011}, metallic conduction in transparent oxides \cite{Hecht2011}, high capacitance in heterostructure forms, large thermopower \cite{Ohta2007}, and the quantum Hall effect \cite{Ezawa2013}. If Kondo lattices can be made 2D, even more fascinating ground states are expected to result.

Numerous attempts to realize heavy fermion systems with low dimensionalities have been reported. The ability to tune the dimensionality of heavy fermion systems to track the evolution of the strength of many-body interactions is particularly desirable. One approach is to insert ``block layers" into a 3D Kondo lattice. This approach has been used to grow bulk crystals of Ce$T$In$_5$ ($T$ = Rh, Co, Ir) \cite{Hegger2000,Petrovic2001,Petrovic2001b}, whose crystal structure comprises alternating layers of CeIn$_ 3$ and $T$In$_2$. According to the band-structure calculations, the $T$In$_2$ block layers are less conductive than the CeIn$_3$ layers \cite{Haga2001}. Therefore, Ce$T$In$_5$ can be treated as a quasi-two-dimensional system. However, the largely corrugated Fermi surface sheets \cite{Settai2007}, 3D-like antiferromagnetic (AFM) fluctuations in the normal state \cite{Zheng2001}, and small anisotropy of upper critical fields \cite{Ida2008} all indicate that the electronic, magnetic, and superconducting properties of Ce$T$In$_5$ are essentially 3D rather than 2D. Another example of 2D Kondo systems is the bilayer 2D films of $^3$He fluid \cite{Neumann2007}, where the mass enhancement is observed near a QCP. Here the control parameter is the $^3$He density of the second layer; thus, this example may involve different physics from $f$-electron-based heavy fermions. Therefore, systematic dimensionality tuning of heavy fermion systems is still lacking. 

Recently, advances in thin-film growth technology have led to the fabrication of artificial 2D structures with controlled atomic-layer thicknesses, providing a unique opportunity to explore novel phenomena in low-dimensional systems with unprecedented control. For example, 2D electron gas \cite{Ohtomo2004}, magnetism \cite{Brinkman2007}, and 2D superconductivity \cite{Reyren2007} have been observed at the interface between insulating nonmagnetic oxides LaAlO$_3$ and SrTiO$_3$; these effects are believed to be closely associated with the interface electronic structure and film thickness. In another example, several peculiar features have been observed in Hubbard superlattices of the AFM Mott--Hubbard insulator LaVO$_3$ and the paramagnetic metal SrVO$_3$; these features include a metal-insulator transition \cite{Sheets2007}, room-temperature ferromagnetism with an insulating interface \cite{Luders2009}, and a resistivity downturn resembling a Kondo coherent state \cite{David2011}, among others. 
Thus, a wide variety of nanometric thin films and superlattices displays notable features arising from their low dimensionality and their interfacial effects. 

Very recently, a state-of-the-art molecular beam epitaxy (MBE) technique has been developed to reduce the dimensionality of the heavy electrons by fabricating multilayers of Ce-based heavy fermion materials sandwiched between conventional metals. In this report, we first examine the heavy fermion properties of bulk CeIn$_3$ and CeCoIn$_5$ to establish a background for later discussions of ``Kondo superlattices". We next discuss the artificial superlattices CeIn$_3$/LaIn$_3$ \cite{Shishido2010}, which undergo a quantum phase transition tuned by the dimensionality of the Kondo lattice, and superconducting Kondo superlattices CeCoIn$_5$/YbCoIn$_5$ \cite{Mizukami2011} in which the superconducting and magnetic properties can be substantially modified by tweaking the degree of spatial inversion symmetry breaking \cite{Goh2012,Shimozawa2014,Yamanaka2015} achieved through a careful design and fabrication of superlattices with desirable layer thicknesses.

\section{Heavy fermion compounds Ce$_n T_m$In$_{3n+2m}$}
The intensively studied heavy fermion compounds CeIn$_3$ and CeCoIn$_5$ belong to the series of compounds with general stoichiometry Ce$_n T_m$In$_{3n+2m}$ ($T$ is a transition-metal element, $n = 1,2$, or $\infty$, and $m = 0, 1$, or $2$) \cite{Thompson2012}. These compounds, which are built of alternating $n$ layers of CeIn$_3$ and $m$ layers of $T$In$_2$ stacked along the tetragonal $c$-axis, are characterized by the interplay of AFM order, unconventional superconductivity, and quantum criticality. For example, CeIn$_3$ \cite{Mathur1998}, CeRhIn$_5$ \cite{Hegger2000}, and CePt$_2$In$_7$ \cite{Bauer2010} are AFM at ambient pressure. When hydrostatic pressure is applied, the AFM order is suppressed and the superconductivity emerges near the QCP. In other examples--specially, CeCoIn$_5$ \cite{Petrovic2001}, CeIrIn$_5$ \cite{Petrovic2001b}, and Ce$_2$PdIn$_8$ \cite{Kaczorowski2009}--an unconventional superconducting state develops from an anomalous normal state that displays breakdown of Landau Fermi liquid behavior. Thus, the Ce$_n T_m$In$_{3n+2m}$ series appears to be an excellent playground for investigating magnetism, superconductivity, and quantum criticality in heavy fermion systems. In this report, CeIn$_3$ and CeCoIn$_5$ are used to study the effect of dimensionality reduction on the antiferromagnetism and unconventional superconductivity in heavy fermion systems.

\subsection{Antiferromagnetic heavy fermion CeIn$_3$}
CeIn$_3$ has a cubic AuCu$_3$-type crystal structure, the simplest crystal structure among the Ce$_n T_m$In$_{3n+2m}$ family. At ambient pressure, this compound undergoes an AFM transition at a N$\acute{\rm e}$el temperature $T_{\rm N}=10$ K, with an ordered moment of $\sim0.5\mu_{\rm B}$/Ce, and the commensurate wavevector ${\bm Q}=(1/2,1/2,1/2)$ \cite{Lawrence1980,Benoit1980}. The 3D AFM state is suppressed monotonically under applied pressure, and a superconducting phase with a maximum $T_c\approx0.2$ K appears near a critical pressure of $P^{\ast}\approx2.5$ GPa \cite{Mathur1998,Knebel2001}. Around a putative AFM QCP, the coexistence of AFM order and superconductivity has been reported \cite{Kawasaki2004}, which implies that the superconductivity may be mediated by magnetic interactions rather than by phonons. Above the superconducting dome, the resistivity strongly deviates from the Fermi liquid behavior, $\rho(T)\sim T^{\alpha}$ with $\alpha=2$, and $\alpha$ approaches 1.6 at the critical pressure where $T_{\rm N}\rightarrow0$. This exponent is similar to the expected value of $3/2$ in the presence of scattering by 3D AFM fluctuations \cite{Moriya2000}, indicating the existence of 3D AFM QCP. Thus, CeIn$_3$ appears to be an ideal candidate for studying the relation between magnetism, unconventional superconductivity, and quantum criticality. Furthermore, the relatively simple electronic and magnetic structures of CeIn$_3$ make this compound particularly appealing.

\subsection{Heavy fermion superconductor CeCoIn$_5$}
The heavy fermion compound CeCoIn$_5$ exhibits superconductivity at 2.3 K, which is the highest transition temperature among the Ce-based heavy fermion systems. CeCoIn$_5$ has attracted extensive interest because of its anomalous normal state transport and superconducting properties, such as non-Fermi liquid (NFL) behavior, QCP, $d_{x^2-y^2}$ superconducting gap symmetry, strong Pauli effect, possible Fulde--Ferrell--Larkin--Ovchinnikov (FFLO) state, and unusual coexistence of superconductivity and magnetic order at low temperatures and high fields. In the following section, we describe the physical properties of CeCoIn$_5$ in detail.

\begin{table*}[t]
\centering
\renewcommand\tabcolsep{8.5pt}
{\caption{Temperature dependences of the physical properties in both Fermi liquid and CeCoIn$_5$. The Landau Fermi liquid theory has been reported to break down in the case of CeCoIn$_5$.}
\label{exponents}
\vspace{1mm}
\newlength{\myheight}
\setlength{\myheight}{0.5cm}
\newlength{\myheighta}
\setlength{\myheighta}{1cm}
\begin{tabular}{cccccccc}
\hline
  & $\rho$ & $C/T$ & $\chi$ (${\bm H}\parallel c$) & $R_{\rm H}$ & $1/T_1$ & $\beta$/$T$\tabularnewline
\hline \hline
\parbox[c][\myheight][c] {0cm}{} Fermi liquid & $T^2$ & const. & const. & const. & $T$ & const.\tabularnewline
\hline
\parbox[c][\myheighta][c] {0cm}{} CeCoIn$_5$ & $T$ (Ref.\,\onlinecite{Sidorov2002}) & -ln $T$ (Ref.\,\onlinecite{Petrovic2001}) & \raisebox{-0.7em}{\shortstack{$T^{-\alpha}$, $(\alpha< 1)$ \\ (Ref.\,\onlinecite{Tayama2002})}} & $1$/$T$ (Ref.\,\onlinecite{Nakajima2004}) & $T^{1/4}$ (Ref.\,\onlinecite{Kohori2001}) & $1$/$T$ (Ref.\,\onlinecite{Zaum2011}) \tabularnewline
\hline
\end{tabular}
}
\end{table*}

\subsubsection{3D electronic, magnetic, and superconducting properties}
CeCoIn$_5$ has a tetragonal HoCoGa$_5$-type crystal structure comprising alternating layers of CeIn$_3$ and CoIn$_2$ stacked along the $c$-axis. Such structural arrangement might imply that CeCoIn$_5$ is a 2D variant of CeIn$_3$. However, the electronic, magnetic, and superconducting natures of CeCoIn$_5$ are observed to be 3D instead of purely 2D. 

First, the result of de Haas--van Alphen (dHvA) measurements suggest that the main Fermi surface sheet with the cyclotron mass $m^\ast_{\rm c} = 87m_0$ is highly corrugated \cite{Settai2007}, which differs substantially from the cylindrical Fermi surface expected for 2D systems. This result indicates that the electronic structure is, if anything, 3D rather than 2D. Second, inelastic neutron scattering measurements detected the presence of magnetic fluctuations at the commensurate wavevector ${\bm Q}=(1/2,1/2,1/2)$ \cite{Stock2008}. Moreover, the nuclear spin relaxation rate exhibits a temperature dependence of the form $1/T_1\sim T^{z}$ with $z=1/4$ \cite{Kawasaki2003}, which is intermediate between $z=0$ and $z=1/2$ for 2D- and 3D-AFM instability, respectively. Therefore, both experimental results provide evidence that rather isotropic AFM fluctuations develop in CeCoIn$_5$. Third, anisotropy of the upper critical field is small ($H_{\rm c2}^{H\parallel ab}/H_{\rm c2}^{H\perp ab}\sim 2.4$) \cite{Ida2008} compared to the case of 2D superconducting systems, which means that anisotropic 3D superconductivity is realized in CeCoIn$_5$. Thus, considering all these factors, CeCoIn$_5$ should be treated as an ``anisotropic" 3D heavy fermion compound at best.

\subsubsection{Non-Fermi liquid behavior}
In conventional metals, the low-energy physics of the system can be described by Fermi liquid theory. As per this theory, physical properties such as electrical resistivity, specific heat, magnetic susceptibility, Hall coefficient, nuclear spin relaxation rate, and thermal expansion coefficient exhibit characteristic temperature dependence; {\it i.e.}, (1) the electrical resistivity $\rho$ varies as $T^2$; (2) the specific heat divided by temperature ($C$/$T$) is a constant; (3) the magnetic susceptibility $\chi$ tends to a constant Pauli-like value; (4) the Hall coefficient $R_{\rm H}$ is independent of temperature; (5) the nuclear spin relaxation rate $1$/$T_1$ is proportional to $T$; and (6) the volume thermal expansion coefficient ($\beta$/$T$) is constant. However, standard Fermi liquid theory has been reported to break down in the case of CeCoIn$_5$.

This non-Fermi liquid (NFL) behavior commonly manifests as a qualitative deviation of the temperature dependences of the physical properties from conventional Fermi liquid behavior. In fact, NFL behavior in CeCoIn$_5$ can be observed in many ways, as evident in Table \ref{exponents}. For example, the resistivity shows $\rho\sim T$ \cite{Sidorov2002}, the specific heat exhibits $C$/$T\sim -\ln{T}$ \cite{Petrovic2001}, the magnetic susceptibility $\chi$ diverges with decreasing temperature \cite{Tayama2002}, the amplitude of $R_{\rm H}$ increases drastically as $1$/$T$ \cite{Nakajima2004}, the nuclear spin relaxation rate follows $1$/$T_1\sim T^{1/4}$ \cite{Kohori2001}, and the volume thermal expansion coefficient $\beta$/$T$ diverges as $1$/$T$ \cite{Zaum2011}. Thus, the temperature dependences of $\rho$, $C$/$T$, $\chi$,  $R_{\rm H}$, $1$/$T_1$, and $\beta$/$T$ are qualitatively different from those of a Fermi liquid (see Table \ref{exponents}). This result is attributed to strong fluctuations associated with the AFM QCP, as described in the following section.

\subsubsection{Quantum critical point (QCP)}
The NFL behaviors of CeCoIn$_5$ at zero field and ambient pressure imply the presence of a QCP nearby. Key information about the QCP is also provided by both (a) the pressure and (b) magnetic field dependences of the physical properties. Until now, many experiments performed at high-pressure and/or high-field suggest that CeCoIn$_5$ is very close to an AFM QCP. The following are the details of these experiments.

\begin{figure*}[t]
\begin{center}
\includegraphics[width=0.9\linewidth]{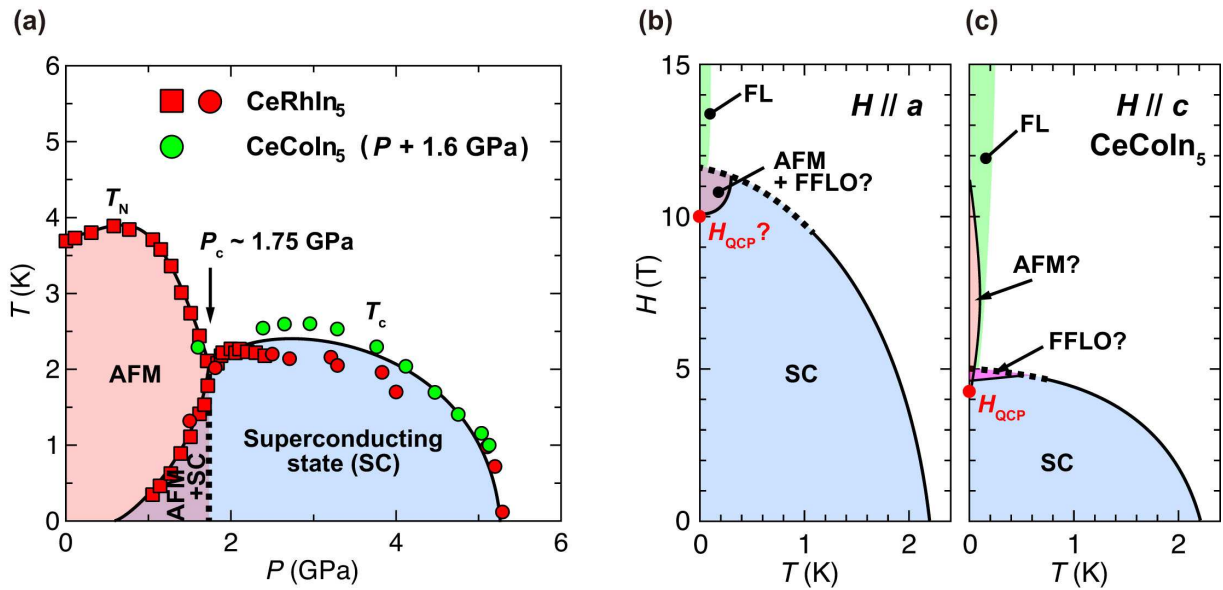}
\caption{Schematics of the phase diagrams. (a) Schematic $P$--$T$ phase diagram at zero magnetic field for CeRhIn$_5$ and CeCoIn$_5$. The pressure axis for CeCoIn$_5$ has been shifted by 1.6 GPa to reflect its smaller unit-cell volume relative to that of CeRhIn$_5$. The experimental data are taken from the specific heat measurements of CeRhIn$_5$ \cite{TPark2008} (red squares) and from the resistivity results for CeRhIn$_5$ \cite{Shishido2002} (red circles) and CeCoIn$_5$ \cite{Sidorov2002} (green circles). $T_{\rm N}$ and $T_{\rm c}$ represent the N$\acute{\rm e}$el and superconducting transition temperatures, respectively. The dashed and solid lines represent the first-order and second-order phase boundaries, respectively. The AFM order coexists with superconductivity below $P_{\rm c}$, whereas it is destroyed above $P_{\rm c}$. (b),(c) Schematic $H$--$T$ phase diagrams for CeCoIn$_5$ at ambient pressure for (b) ${\bm H}\parallel a$ and (c) ${\bm H}\parallel c$. $H_{\rm QCP}$ are slightly smaller than $H_{\rm c2}$ for both field directions. A field-induced AFM phase (Q-phase) was identified for ${\bm H}\parallel a$, whereas no signature of an AFM state was observed for ${\bm H}\parallel c$.}
\label{Phase diagram 115}
\end{center}
\vspace{-5mm}
\end{figure*}

\vspace{5mm}
\begin{center}
{\noindent}{(a) Pressure dependence}
\end{center}

According to the resistivity measurements, the NFL behavior is suppressed and a Fermi liquid state with $\rho=\rho_0 + AT^2$ is recovered with increasing pressure \cite{Sidorov2002}. At a critical pressure of $P^{\ast}\approx1.6$ GPa, a rapid change in the residual resistivity $\rho_0$ and the coefficient $A$ is observed. Because both $\rho_0$ and $A$ are expected to be enhanced upon approaching the QCP because of the strong fluctuations, this change indicates that CeCoIn$_5$ is located very close to the QCP at ambient pressure. In fact, Cd doping is known to induce an AFM order in CeCoIn$_5$ \cite{Pham2006}, in which Cd doping appears to mimic the effect of negative pressure, although the changes in carrier density and disorder are also caused by Cd substitution. Moreover, the unit-cell volume of CeCoIn$_5$ at ambient pressure has been reported to be equivalent to that of the isostructural AFM compound CeRhIn$_5$ at $P=1.6$ GPa \cite{Kumar2004}, which is smaller than but similar to $P_{\rm c} \sim 1.75$ GPa, where $T_{\rm N} \rightarrow 0$. In the zero-field $P$--$T$ phase diagram (Fig. \ref{Phase diagram 115}(a)), the shape of the superconducting dome in CeCoIn$_5$ also coincides with that in the isostructural AFM compound CeRhIn$_5$ \cite{Thompson2006} when CeCoIn$_5$ at ambient pressure is assumed to be equivalent to CeRhIn$_5$ at $1.6$ GPa.
These results reinforce our speculation that the AFM QCP in CeCoIn$_5$ may lie at an inaccessible, slightly negative pressure.

This assumption is also supported by the results of the specific heat \cite{Sparn2002} and dHvA measurements \cite{Shishido2003}, which suggest that CeCoIn$_5$ is pushed away from the QCP under pressure because the effective mass decreases with increasing pressure. Furthermore, the Hall coefficient measurements indicate that the enhancement of $|R_{\rm H}|$ at low temperatures is drastically suppressed at high pressures \cite{Nakajima2007}, implying the suppression of the fluctuations associated with the AFM QCP. Such a change in the magnetic fluctuation is directly observed by the nuclear quadrupole resonance (NQR). In CeCoIn$_5$, the nuclear spin relaxation rate divided by the temperature, $1$/$T_1T$, is well described by $1$/$T_1T\propto (T+\theta_W)^{-3/4}$. As per spin fluctuation theory, the Weiss temperature $\theta_W$ is regarded as a scaling parameter that benchmarks the distance to the AFM QCP from the phase-space location of the system. At ambient pressure, $\theta_W = 0.6$ K has been reported; however, $\theta_W$ increases with pressure \cite{Yashima2004}, which again indicates the existence of a pressure-induced AFM QCP close to ambient pressure in CeCoIn$_5$.

\vspace{5mm}
\begin{center}
{\noindent}{(b) Magnetic field dependence}
\end{center}

The low-temperature resistivity measurements of the normal state above the upper critical fields, $H_{\rm c2}\sim 11.8$ T for ${\bm H}\parallel ab$ \cite{Ronning2005} and $H_{\rm c2}\sim 4.95$ T for ${\bm H}\parallel c$ \cite{Paglione2003}, have revealed that a Fermi liquid state with $\rho=\rho_0 + AT^2$ is recovered at high magnetic fields. The prefactor $A$ exhibits divergent behavior toward the critical field $H_{\rm QCP}$ for both ${\bm H}\parallel ab$ and ${\bm H}\parallel c$, which provides strong evidence for the presence of field-induced QCP. The quantum criticality was also detected in the measurements of specific heat \cite{Bianchi2003,Bianchi2003b}, magnetization \cite{Tayama2002}, thermal transport \cite{Paglione2014}, Hall effect \cite{Singh2007}, and thermoelectric coefficients \cite{Izawa2007}. Remarkably, $H_{\rm QCP}$ almost coincides with $H_{\rm c2}$ at ambient pressure, irrespective of field orientation \cite{Ronning2005} or suppression of the superconducting state by Sn doping \cite{Bauer2005}. This result naturally implies that the field-induced quantum criticality might be inherently linked with the superconductivity. However, the $H_{\rm c2}$ transition has first-order character at low temperatures, as described later. Furthermore, the high-pressure resistivity measurements demonstrate that $H_{\rm QCP}$ is suppressed with increasing pressure faster than $H_{\rm c2}$ \cite{Ronning2006}. These studies indicate that the superconducting fluctuations are unlikely to be responsible for the quantum critical behaviors at $H_{\rm QCP}$. 

A comparison between CeCoIn$_5$ and AFM CeRhIn$_5$ provides useful information about the nature of the critical fluctuations at $H_{\rm QCP}$. In CeRhIn$_5$ at zero magnetic field, the AFM order is completely destroyed at the critical field of $P\sim 1.75$ GPa; however, the AFM phase is resurrected in the superconducting state under magnetic fields and is expanded in magnetic fields stronger than $H_{\rm c2}$ \cite{Ida2008,Knebel2011}. This fact suggests that the field-induced quantum criticality in CeCoIn$_5$ is also likely associated with an as yet undetected AFM fluctuation because CeCoIn$_5$ is located close to the AFM QCP at zero field. Indeed, a field-induced AFM phase (Q-phase) was recently found in CeCoIn$_5$ close to $H_{\rm QCP}$ for ${\bm H}\parallel ab$ \cite{Kenzelmann2008}, as shown in Fig. \ref{Phase diagram 115}(b). By contrast, no signature of an AFM state was observed around $H_{\rm QCP}$ for ${\bm H}\parallel c$. The AFM phase may exist at ultra-low temperatures inaccessible by a conventional dilution refrigerator (Fig. \ref{Phase diagram 115}(c)).

\subsubsection{Superconductivity}
The zero-field-cooled magnetic susceptibility and resistivity provide clear evidence that CeCoIn$_5$ exhibits bulk superconductivity at $T_c$ = 2.3 K \cite{Petrovic2001}. The large $\gamma$ = $C$/$T$ = 290 mJ mol$^{-1}$K$^{-2}$ is observed just above $T_c$ \cite{Ikeda2001}, indicating that the quasiparticle effective mass is substantially enhanced at low temperatures. The steep initial slope of the upper critical field near $T_c$ confirms that such heavy quasiparticles form Cooper pairs. Moreover, the jump size of the specific heat at $T_c$ is remarkably large, and its value is $\Delta C$/$\gamma T_c \sim$ 4.5, which suggests that CeCoIn$_5$ exhibits very strong coupling superconductivity (the expected value for a weak coupling superconductor is only $\Delta C$/$\gamma T_c$ = 1.43). 

\subsubsection{Superconducting gap structure}
An unambiguous determination of the pairing symmetry in unconventional superconductors is essential to an understanding their pairing mechanism and the origin of their superconductivity. In CeCoIn$_5$, the temperature dependence of the zero-field heat capacity ($C\propto T^2$) and the residual linear term in the thermal conductivity ($\kappa\propto T$) indicate the presence of line nodes in the superconducting energy gap \cite{Movshovich2001}. Moreover, the spin relaxation rate, $1$/$T_1$ \cite{Kohori2001}, obtained by NQR measurements is approximately described by a $1$/$T_1\propto T^3$ law and no Hebel--Slichter coherence peak is observed at $T_c$, which is significantly different from the behaviors of conventional superconductors. These measurements thus provide strong evidence for unconventional superconductivity with line nodes in the superconducting gap. The penetration depth measurements also support this conclusion, although the low-$T$ exponent deviates from unity \cite{Ormeno2002,Ozcan2003}, which may be attributable to strong fluctuations arising from the QCP \cite{Hashimoto2013}.

In unconventional superconductors, {\it i.e.}, those with non-$s$-wave pairing states, the spin-part of the pairing wavefunction can have two types of pairing symmetry: spin-singlet pairing or spin-triplet pairing. $^{115}$In and $^{59}$Co nuclear magnetic resonance (NMR) Knight shift \cite{Kohori2001,Curro2001} and torque measurements \cite{Xiao2007} reveal the suppression of spin susceptibility below $T_c$, which implies that CeCoIn$_5$ is a singlet superconductor. This result is consistent with the observation of first-order superconducting-to-normal states transition because of the Pauli pair-breaking effect at high magnetic fields and low temperatures. Given the existence of line nodes in the superconducting gap, the superconducting order parameter of CeCoIn$_5$ must possess $d$-wave symmetry. 

For CeCoIn$_5$, these line nodes can assume one of two forms: $d_{x^2-y^2}$ or $d_{xy}$ symmetry. The direction of the nodes has been measured using various techniques. In the thermal conductivity \cite{Izawa2001} and specific heat measurements \cite{An2010}, a four-fold oscillation is clearly observed in the case of a magnetic field rotating in the $ab$ plane. The location of these maximum values is consistent with a $d_{x^2-y^2}$-wave symmetry order parameter. Point-contact spectroscopy \cite{WPark2008}, which has been used to observe the Andreev reflection signals in the normal metal/CeCoIn$_5$ junctions along two crystallographic orientations, [100] and [110], has also led to the conclusion that the gap symmetry is $d_{x^2-y^2}$. Moreover, small-angle neutron scattering experiments have indicated the presence of a square vortex lattice \cite{Bianchi2008}, whose orientation relative to the crystal lattice is consistent with that expected for $d_{x^2-y^2}$ symmetry. Most recently, scanning tunneling microscopy measurements \cite{Allan2013,Zhou2013} have supported these conclusions through measurements of the response of superconductivity to various perturbations via quasiparticle interference and local pair-breaking experiments. Given these results from a wide range of measurements, the nodes in CeCoIn$_5$ are most consistently of $d_{x^2-y^2}$ gap symmetry, which implies that the AFM fluctuations are important for superconductivity. This observation is in qualitative agreement with the results of NMR \cite{Kawasaki2003} and neutron scattering \cite{Stock2008} experiments that indicated the occurrence of anisotropic spin fluctuations.

\subsubsection{Pauli-limited superconductivity, FFLO state, and Q-phase}
Magnetic fields destroy superconductivity through two mechanisms: the orbital pair-breaking and Pauli pair-breaking effects. In the former case, Cooper pairs are destroyed by the effect of Lorentz force, whereas, in the latter case, superconductivity is destroyed by the Zeeman splitting of spin-singlet Cooper pairs. In the case of a superconductor with strong Pauli effect, the superconducting transition has been predicted to change from second to first order below $T/T_c = 0.55$ \cite{Sarma1963,Maki1964}; however, for almost 40 years, no conclusive experimental evidence supporting this conjecture had been reported. Recently, the first-order phase transition was, as far as we know, first observed in thermal conductivity measurements of CeCoIn$_5$ \cite{Izawa2001}, indicating that superconductivity is dominated by Pauli paramagnetism. This conclusion was also confirmed by the step-like features of the specific heat \cite{Radovan2003,Bianchi2003}, magnetization \cite{Tayama2002,Okazaki2007}, magnetocaloric \cite{Tokiwa2012}, and thermal expansion measurements \cite{Takeuchi2002,Oeschler2003}.

In CeCoIn$_5$ with a strong Pauli effect, the Maki parameters, $\alpha_{\rm M}\sim 4.6$ for $\bm{H}\parallel ab$ and $\alpha_{\rm M}\sim 3.6$ for $\bm{H}\parallel c$, are substantially greater than the minimum required value of $\alpha_{\rm M} = 1.8$ for realizing the spatially inhomogeneous FFLO state \cite{Fulde1964,Larkin1965}, where $\alpha_{\rm M}$ represents the ratio between the orbital and Pauli limiting critical fields, $\alpha_{\rm M}=\sqrt{2}H_{c2\perp}^{\rm orb}(0)/H_{c2\perp}^P(0)$ \cite{James1969}. Additionally, CeCoIn$_5$ crystals can be made very pure; \cite{Kasahara2005} thus, the stringent requirement of ultrahigh purity for the stabilization of the FFLO state is comfortably satisfied. Therefore, the FFLO state is expected to be realized in CeCoIn$_5$. Thus far, signatures of a new superconducting phase have been observed by several experiments, including heat capacity \cite{Radovan2003,Bianchi2003}, ultrasound \cite{Watanabe2004}, penetration depth \cite{Martin2005}, and thermal conductivity measurements \cite{Capan2004} at low temperatures in the vicinity of the upper critical field $H_{\rm c2}$ for both $\bm{H}\parallel ab$ and $\bm{H}\parallel c$. This new phase exhibits properties consistent with the formation of the FFLO state; hence, after $\sim$40 years of intensive research, CeCoIn$_5$ could indeed be a material that hosts the long sought FFLO phase.

$^{115}$In NMR measurements \cite{Kakuyanagi2005,Kumagai2006} provide direct microscopic information about the low-temperature, high-magnetic-field phase near $H_{c2}$ in CeCoIn$_5$: (a) at fields below $H_{c2}$, a new resonance peak appears with a small but finite intensity at higher frequencies, and (b) its frequency coincides well with the resonance frequency obtained in the normal state. The first feature implies that the weak peak is unlikely to be generated by AFM ordering, because, if AFM order develops, the alternating hyperfine fields would produce two inequivalent In(1) sites, which generates two resonance peaks with equal intensities. The second feature suggests that the higher resonance peak is a manifestation of a novel normal-state quasiparticle structure in the high-field, low-temperature region of the phase diagram. These observations point to the possibility of an FFLO state in CeCoIn$_5$.

Investigations of the presumed inhomogeneous superconducting state are further complicated by the recent observation of a field-induced long-range incommensurate AFM phase (Q-phase) detected by NMR/NQR \cite{Young2007,Koutroulakis2010} and neutron scattering \cite{Kenzelmann2008,Kenzelmann2010}. This magnetically ordered state emerges only in the superconducting mixed state and is observed in the same high-field, low-temperature region of the phase diagram as the region where the FFLO state was proposed to exist. The ordering wavevector ${\bm Q}=(0.45,0.45,1/2)$ is the same when $\bm{H}$ is applied parallel to the [100] and [110] crystallographic directions \cite{Gerber2014}, which is in contradiction with the predictions for the FFLO state. Several theoretical explanations have been proposed to account for this characteristic relationship between superconductivity and magnetism. Some hypotheses suggest that the Q-phase is independent of a possible FFLO state \cite{Agterberg2009} and, consequently, coexists with the nodal, $d_{x^2-y^2}$ wave-gap symmetry, whereas others emphasize that the Q-phase is allowed to be stabilized because of the presence of the FFLO state \cite{Yanase2009}. Thus, whether this Q-phase replaces or coexists with a non-standard FFLO state remains unclear; however, recent NMR experiments have at least revealed a quasiparticle density of states consistent with coexisting FFLO nodal planes and long-range AFM order \cite{Kumagai2011}.

\section{Experimental methods}
\begin{figure}[t]
\begin{center}
\includegraphics[width=0.95\linewidth]{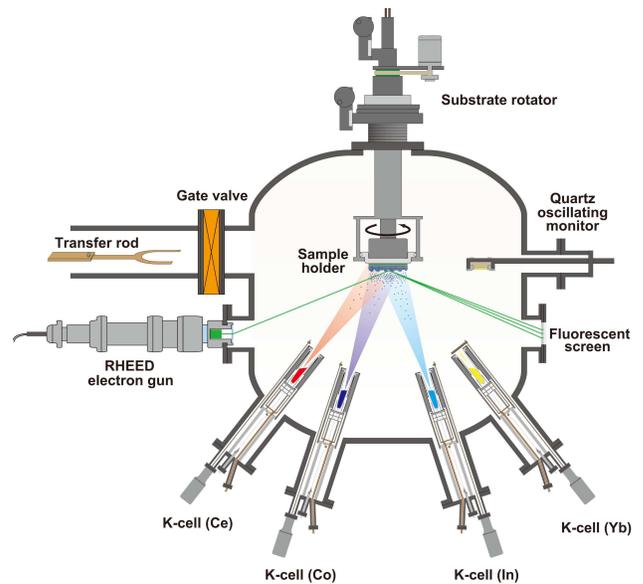}
\caption{Schematic of an MBE system for fabrication of Kondo superlattices. The molecular beams are generated by heating solid materials placed in Knudsen cells (K-cells), which are then directed to the crystalline substrate surface. Due to the very large free mean path under ultra-high vacuum conditions, very few collisions occur between the particles. The substrate is placed on a sample holder, which is heated to the necessary temperature by resistive heaters. The deposition rate of each element is precisely controlled by a quartz oscillator thickness monitor. The MBE system is equipped with the RHEED, which is used for monitoring the morphology and roughness of the sample surface. The diffraction patterns are instantaneously displayed on the fluorescent screen located on the diametrically opposite side of the vacuum chamber from the electron gun. In this configuration, the growing surface can be continuously monitored during the sample growth.}
\label{MBE}
\end{center}
\vspace{-5mm}
\end{figure}

Our group has pioneered an MBE technique (Fig. \ref{MBE}) to fabricate artificial heavy fermion superlattices in a controllable fashion using layer-by-layer epitaxial growth of Ce-based materials. Under certain conditions, such Kondo superlattices can essentially be regarded as a 2D Kondo lattice. Recently, we have also developed a mothod to tune the degree of inversion symmetry breaking (ISB) by designing new types of superlattices. In this section, we first discuss the advantages of our MBE technology for the fabrication of Kondo superlattices. We then discuss the conditions for the realization of a 2D Kondo lattice. We also introduce three types of superconducting Kondo superlattices that we have fabricated thus far: AB-, ABAB$^{\prime}$-, and ABC-type superlattices (Fig. \ref{Kondo_SL}).

\subsection{Molecular beam epitaxy (MBE) systems}
MBE is a versatile technique for epitaxial growth by impinging molecular beams of atoms onto a heated substrate under ultra-high-vacuum conditions  ($\sim10^{-7}$ Pa). A schematic of the basic evaporation process in the MBE chamber is shown in Fig. \ref{MBE}. In the following sections, we briefly introduce the main features of the MBE method and our ultimate choice of MgF$_2$ as the substrate material, which are key factors for the successful fabrication of multilayers of CeIn$_3$ and CeCoIn$_5$.

\subsubsection{Features of MBE technology}
The characteristic features of the MBE method are described as follows. (1) The MBE is essentially a refined ultra-high-vacuum evaporation method, which helps to prevent contamination of the surface and oxidation of elements such as Ce; thus, high-quality thin films of Ce-based compounds can be grown using MBE. (2) MBE enables a slow growth rate of 0.01--0.02 nm/s that permits very precise control of layer thickness. Consequently, abrupt material interfaces can be achieved, enabling the fabrication of heterostructures such as superlattices. (3) The typical pressure in the MBE chamber is maintained at $\leq10^{-7}$ Pa during the fabrication of thin films, which enables powerful diagnostic techniques such as reflection high-energy electron diffraction (RHEED) for {\it in situ} monitoring of thin films' growth without the complication of surface degradation. Thus, MBE technology is well suited for achieving our goal of designing Kondo superlattices for dimensionality tuning of strongly correlated electron systems.

\subsubsection{MgF$_2$ substrate}
Several significant factors affect the growth of thin films. Among them, lattice matching between the substrate and the thin film is considered one of the most critical factors for epitaxial growth. Magnesium fluoride (MgF$_2$) has a rutile-type tetragonal structure with a lattice parameter $a=0.462$ nm, which matches the lattice parameters $a=0.468$ nm and $a=0.461$ nm for CeIn$_3$ and CeCoIn$_5$, respectively. Furthermore, because MgF$_2$ does not contain oxygen, the oxidation of Ce compounds during the growth can be avoided. Thus, single-crystal MgF$_2$ is a suitable substrate material to support the epitaxial growth of CeIn$_3$ and CeCoIn$_5$ thin films. 

The preparation of a clean MgF$_2$ substrate is another important factor in MBE deposition. We first prepared an atomically smooth MgF$_2$ surface using chemical mechanical polishing techniques (Crystal Base Ltd.). We next loaded the substrate into a high-vacuum sample transfer area after cleaning it with compressed air and then transferred it to the MBE growth chamber, where it was baked at 750$^{\circ}$C for 2 h to remove most of the contaminants from the substrate surface. Finally, the substrate temperature was reduced to approximately 550$^{\circ}$C, which is the optimal growth temperature. After the substrate was baked, a clear streak pattern of the MgF$_2$ substrate appeared in the RHEED image, which indicates that the substrate had a very smooth and crystalline surface.

\begin{figure*}[t]
\begin{center}
\includegraphics[width=0.9\linewidth]{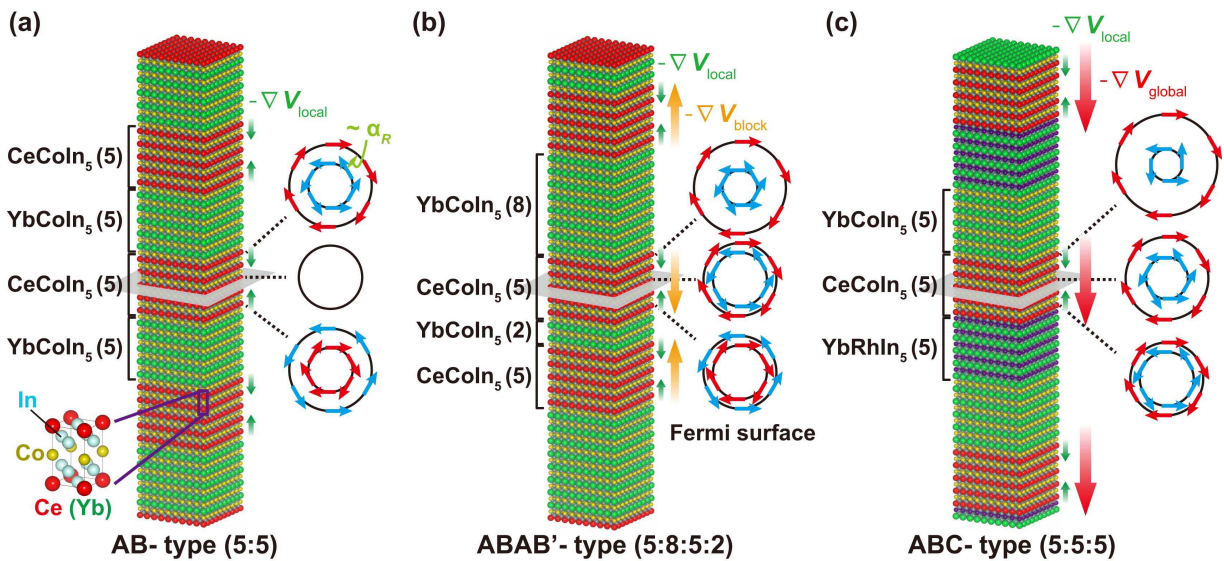}
\caption{Schematics of three types of superconducting Kondo superlattices. (a) AB-type: Superlattice with alternating layers of $5$-UCT CeCoIn$_5$ and $5$-UCT YbCoIn$_5$, ($n$:$m$) = ($5$:$5$). The middle CeCoIn$_5$ layer in a given CeCoIn$_5$ BL indicated by the gray plane is a mirror plane. The green small arrows represent the asymmetric potential gradient associated with the local ISB, $-\nabla V_{\rm local}$. The Rashba splitting, $\alpha_R$, occurs at the interface CeCoIn$_5$ layers neighboring the YbCoIn$_5$ layers because of the local ISB. The spin direction is rotated in the $ab$ plane and is opposite between the top and bottom CeCoIn$_5$ layers. (b) ABAB$^{\prime}$-type: $5$-UCT CeCoIn$_5$ BLs are sandwiched by $8$- and $2$-UCT YbCoIn$_5$ layers, ($n$:$m$:$n$:$m^{\prime}$) = ($5$:$8$:$5$:$2$). The middle CeCoIn$_5$ layer (gray plane) is not a mirror plane. The orange medium arrows represent the asymmetric potential gradient associated with the YbCoIn$_5$ layer thickness modulation $-\nabla V_{\rm block}$, which is oriented in the opposite direction in the neighboring CeCoIn$_5$ BLs. (c) ABC-type: Tricolor superlattice consisting of $5$-UCT CeCoIn$_5$, $5$-UCT YbCoIn$_5$, and $5$-UCT YbRhIn$_5$, ($n$:$m$:$l$)=($5$:$5$:$5$). All layers, including the middle CeCoIn$_5$ layer (gray plane), are not mirror planes. The asymmetric potential gradient (red large arrows) associated with global ISB is oriented in the same direction for all CeCoIn$_5$ BLs.}
\label{Kondo_SL}
\end{center}
\vspace{-5mm}
\end{figure*}

\subsection{2D Kondo lattice}
To realize a 2D Kondo lattice, we fabricated the superlattices with alternating block layers (BLs) of $n$ unit-cell-thick (UCT) Ce compound and $m$-UCT non-magnetic compound, such as CeIn$_3$($n$)/LaIn$_3$($m$), CeCoIn$_5$($n$)/YbCoIn$_5$($m$), and CeRhIn$_5$($n$)/YbRhIn$_5$($m$) superlattices. In these superlattices, each Ce BL can be treated as an isolated 2D Kondo lattice when the following two conditions are satisfied: (i) Ce BLs are magnetically isolated from each other when non-magnetic BLs with $m \gtrsim 2$ are inserted; this arrangement is achievable because the interlayer magnetic interaction mediated through the RKKY interaction between the Ce ions in different BLs decreases to less than $1/(m+1)^3$ of that between the neighboring Ce ions within the same layer \cite{Peters2013}. (ii) The dimensionality of each Ce BL is reduced from three to two by forming the Ce BL with a few UCT ($n\lesssim 5$). Thus, a 2D Kondo lattice can be achieved using a Kondo superlattice.


\subsection{Superconducting Kondo superlattices}
One of our primary interests is to develop a scheme to control the degree of ISB through an appropriate design of Kondo superlattices. We have succeeded in designing and fabricating several types of superconducting superlattices, specially, AB-, ABAB$^{\prime}$-, and ABC-type superlattices. These superlattices all contain heavy fermion superconductor CeCoIn$_5$ and conventional metals YbCoIn$_5$ and/or YbRhIn$_5$. 

The AB-type superlattice is a conventional superlattice with alternating BLs of $n$-UCT CeCoIn$_5$ and $m$-UCT YbCoIn$_5$ or YbRhIn$_5$, forming an ($n$:$m$) heterostructure. Spatial inversion symmetry is locally broken at the top and the bottom ``interface" CeCoIn$_5$ layers in the immediate proximity to the Yb BLs, which induces an asymmetric potential gradient associated with the local ISB, $-\nabla V_{\rm local}$ \cite{Maruyama2012,Goh2012}, as indicated by the green (small) arrows in Fig. \ref{Kondo_SL}(a). In this case, the Fermi surface is split into two sheets with different spin structures: the spin direction is tilted into the plane, rotating clockwise on one sheet and anticlockwise on the other. Because the fraction of noncentrosymmetric interface layers increases rapidly with decreasing thickness of Ce BLs, the effect of this ``{\it local}" ISB is expected to be more pronounced with decreasing $n$.

The ABAB$^{\prime}$-type superlattice is a modulated superlattice consisting of $n$-UCT CeCoIn$_5$ sandwiched by $m$- and $m^{\prime}$-UCT YbCoIn$_5$ or YbRhIn$_5$, denoted as an ($n$:$m$:$n$:$m^{\prime}$) superlattice. In this superlattice, inversion symmetry is disrupted in all of Ce planes through the thickness modulation of the Yb BLs \cite{Shimozawa2014}, in addition to the local ISB. The asymmetric potential gradient associated with the Yb-BL thickness modulation, $-\nabla V_{\rm block}$, is oriented in the opposite direction in the neighboring CeCoIn$_5$ BLs, as shown by the orange (medium) arrows in Fig. \ref{Kondo_SL}(b). We note that mirror planes are present in Yb BLs even in the ABAB$^{\prime}$-type superlattice; however, we here focus mainly on ISB in the Ce-planes. We expect that the degree of this ``{\it block layer}" ISB can be enhanced with increasing $|m-m'|$, which represents the degree of thickness modulation of Yb BLs.  

The ABC-type superlattice is a tricolor superlattice composed of $n$-UCT CeCoIn$_5$, $m$-UCT YbCoIn$_5$, and $l$-UCT YbRhIn$_5$, forming an ($n$:$m$:$l$) $c$-axis-oriented superlattice structure. In this case, we find immediately that inversion symmetry is globally broken along the $c$-axis, as well as the local ISB. The asymmetric potential gradient associated with global ISB, $-\nabla V_{\rm global}$, is oriented in the same direction for all CeCoIn$_5$ BLs, as shown by the red (large) arrows in Fig. \ref{Kondo_SL}(c), in stark contrast to the case of block-layer ISB. We anticipate that the degree of this ``{\it global}" ISB can be tuned by changing the ratio between the thickness of YbCoIn$_5$ and that of YbRhIn$_5$, $m/l$.

Thus, through fabrication of superlattices with precise control of material composition and layer thicknesses, three types of ISB can be introduced: local, block-layer, and global ISB. In this report, we discuss the effect of local and/or block-layer ISBs on the superconductivity and magnetic properties.

\section{Tuning the quantum criticality in CeIn$_3$/LaIn$_3$ superlattices}
We first examine how the dimensionality of the Kondo lattice affects the AFM state. In reduced spatial dimensions, magnetic order is often suppressed because of quantum fluctuations. Therefore, the dimensionality can be regarded as a tuning parameter for quantum criticality. Here, to adjust the dimensionality in a controllable manner, we used MBE to grow AB-type superlattices with alternating layers of $n$-UCT CeIn$_3$ and $m$-UCT LaIn$_3$, forming an ($n$:$m$) heterostructure \cite{Shishido2010}.

\begin{figure}[t]
\begin{center}
\vspace{2mm}
\includegraphics[width=0.95\linewidth]{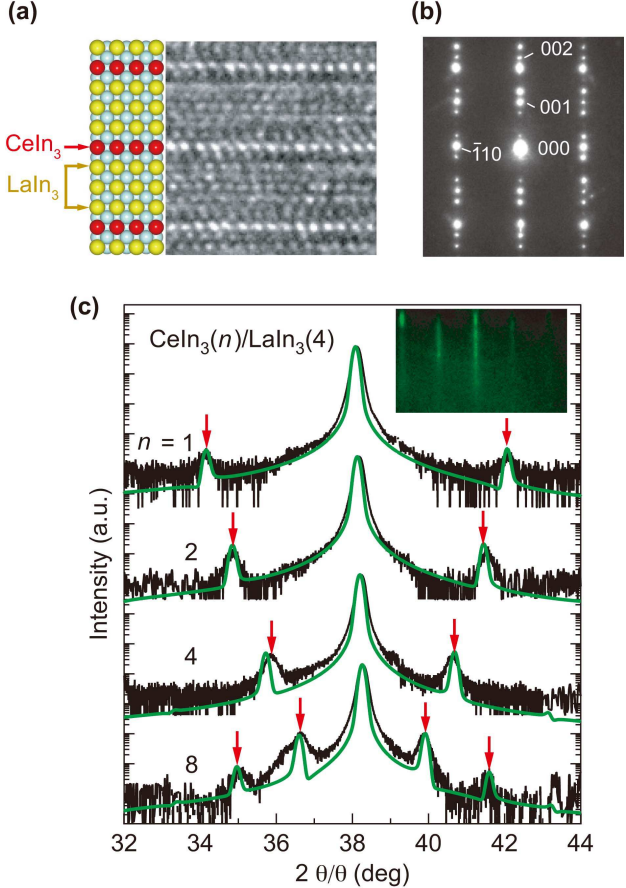}
\caption{Superlattice structures of CeIn$_3$ and LaIn$_3$. (a) Cross-sectional TEM image of the ($n$:$m$)=($1$:$3$) superlattice. The Ce atoms are identified as brighter spots than the La and In atoms. The left image represents the designed ($1$:$3$) superlattice structure. (b) The diffraction pattern of the electron beam incident along the [110] direction. (c) X-ray diffraction patterns showing the 2$\theta$ regions around the (002) peak of the ($n$:$4$) superlattices for $n=1$, $2$, $4$, and $8$. Green lines represent the step model simulations ignoring interface and layer-thickness fluctuations; these simulations reproduce both the intensities and the positions of the satellite peaks (arrows). Inset: streak patterns in the RHEED image of a typical CeIn$_3$/LaIn$_3$ superlattice.}
\label{13SL_characterization}
\end{center}
\vspace{-5mm}
\end{figure}

\subsection{Characterization of CeIn$_3$/LaIn$_3$ superlattices}
A cross-sectional transmission electron microscopy (TEM) image for the ($n$:$m$)=($1$:$3$) superlattice (Fig. \ref{13SL_characterization}(a)), in which the bright spots correspond to Ce atoms, provides direct evidence for the formation of the desired superlattice structure shown on the left hand side of the TEM image. The diffraction pattern of the electron beam incident along the [110] direction shows clear superspots (Fig. \ref{13SL_characterization}(b)), which reflects the long-period stacking structures based on the superlattices. The X-ray diffraction patterns (Fig. \ref{13SL_characterization}(c)) agree well with the step model simulations \cite{Fullerton1992} (green lines), demonstrating the realization of epitaxial superlattice structures, with no discernible interdiffusion even for $n = 1$ cases. A clear streak pattern was also observed in the RHEED image (inset of Fig. \ref{13SL_characterization}(c)), indicating epitaxial growth of each layer with atomic-scale flatness. Similar streak patterns are also observed for other superlattices, which confirms the epitaxial growth with smooth surfaces for all ($n$:$m$) superlattices. These results indicate the successful confinement of $f$-orbital electrons within the 2D space.

\begin{figure*}[t]
\begin{center}
\includegraphics[width=0.8\linewidth]{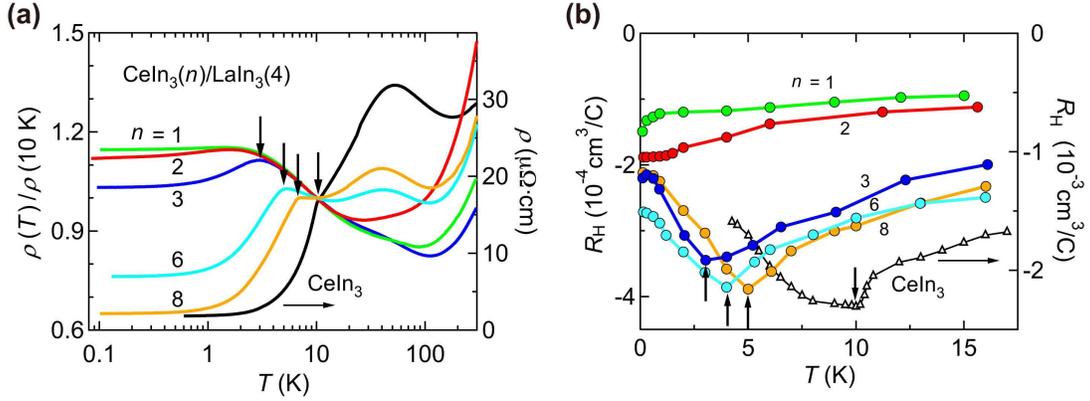}
\vspace{-2mm}
\caption{Transport properties in the artificial superlattice CeIn$_3$/LaIn$_3$. (a) Temperature dependence of the resistivity normalized by its value at 10 K, $\rho (T)$/$\rho$(10 K), in the ($n$:$4$) superlattices for $n=1$ (green), $2$ (red), $3$ (dark blue), $6$ (light blue), and $8$ (yellow). The $\rho(T)$ of the CeIn$_3$ thin film (black line, right axis) coincides well with the results for the bulk material. (b) Temperature dependence of the Hall coefficient $R_{\rm H}$ for the superlattices (left axis) and for bulk CeIn$_3$ (right axis). Both $\rho (T)$ and $R_{\rm H}(T)$ show cusp-like behaviors (arrows) for $n=8$, $6$, and $3$ as well as for CeIn$_3$.}
\label{13SL_Neel}
\end{center}
\vspace{-5mm}
\end{figure*}

\subsection{Suppression of N$\acute{\rm e}$el temperature with the dimensionality reduction}
We investigate the transport properties of the ($n$:$m$) superlattices as a function of CeIn$_3$ layer thickness $n$. As noted above, we set $m=4$ to meet the requirement for realizing the 2D Kondo lattice, in which the interlayer magnetic interaction is substantially reduced. In fact, this 2D feature is manifested by the distinct anisotropy of the magnetoresistance for ($n$:$m$)=(1:4), which is in sharp contrast to the magnetoresistance expected for both CeIn$_3$ and LaIn$_3$, which possess a 3D isotropic structure.

Figures \ref{13SL_Neel}(a) and (b) show the temperature dependence of the in-plane resistivity normalized by its value at 10 K, $\rho (T)$/$\rho$(10 K), and the Hall coefficient $R_{\rm H}$ for $\bm{H}\parallel c$, in each ($n$:$4$) superlattice. For comparison, we also plot $\rho(T)$ for a CeIn$_3$ thin film and $R_{\rm H}(T)$ for bulk CeIn$_3$. In the case of the CeIn$_3$ thin film, $\rho(T)$ reproduces well the results obtained for bulk single crystals; {\it i.e.}, $\rho(T)$ increases because of the Kondo scattering below $\sim 200$ K, followed by a broad hump at $T_{\rm peak} \sim 50$ K and a rapid decrease due to the suppression of the magnetic scattering after a distinct cusp at the N$\acute{\rm e}$el temperature $T_{\rm N} = 10$ K. In the superlattices, the hump structure becomes less pronounced with decreasing $n$, and $\rho(T)$ increases again below $T_{\rm peak}$ because of the interplay of the Kondo interaction with the crystal field effect \cite{Kashiba1986}, where the energy scale is 123 K in CeIn$_3$ \cite{Knafo2003}. At lower temperatures, $\rho(T)$ for $n \geq 3$ is essentially different from those for $n = 2$ and $1$. In the case of $n = 8$, $6$, and $3$, $\rho(T)$ shows a pronounced peak structure, as indicated by arrows in Fig. \ref{13SL_Neel}(a). These peak temperatures coincide well with the temperatures at which $R_{\rm H}(T)$ exhibits cusp-like minima (Fig. \ref{13SL_Neel}(b)) for $n = 8$, $6$, and $3$. Considering that both $\rho (T)$ and $R_{\rm H}$ have the cusp at $T_{\rm N}$ in the case of the CeIn$_3$ thin film, the observed transport anomalies for $n = 8$, $6$, and $3$ appear as a result of AFM ordering. By contrast, for $n=2$ and $1$, $\rho(T)$ does not exhibit a clear peak and $R_{\rm H}(T)$ decreases monotonically with decreasing temperature, without the upturn. These observations indicate that $T_{\rm N}$ is suppressed with decreasing CeIn$_3$ layer thickness, resulting in the disappearance of the AFM order near $n = 2$ (see Fig. \ref{13SL_QCP}(a)). We thus reasonably conclude that the enhanced AFM fluctuations associated with two-dimensionality are responsible for destroying this AFM order, as elaborated in the following section.

\subsection{2D antiferromagnetic QCP}
The vanishing AFM order suggests the existence of a QCP at approximately $n = 2$, although given the limited amount of data, the evidence is not yet conclusive. In the vicinity of the QCP, quantum fluctuations cause the enhancement of the effective mass or the deviation from the Fermi liquid behavior. 

\begin{figure}[bt]
\begin{center}
\vspace{2mm}
\includegraphics[width=0.9\linewidth]{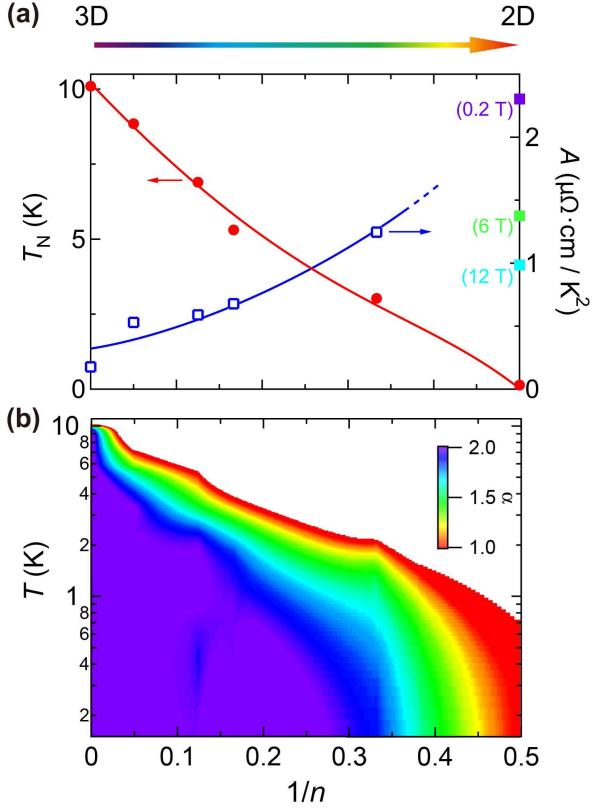}
\vspace{-3mm}
\caption{Temperature versus dimensionality phase diagrams. (a) $T_{\rm N}$ (red circles) and the Fermi liquid coefficient $A$ (open blue squares) as function of 1/$n$. For $n = 2$, the $A$ values are determined from $\rho(T)$ under magnetic fields (solid squares). The solid and dashed lines are visual guides. (b) Temperature and layer thickness evolution of the exponent $\alpha$ derived from the expression $\rho(T) = \rho_0 + AT^{\alpha}$.}
\label{13SL_QCP}
\end{center}
\vspace{-5mm}
\end{figure}

We focus first on the behavior of the effective mass. Away from the QCP, the resistivity follows the Fermi liquid behavior, $\rho(T) = \rho_0 + AT^{\alpha}$ with $\alpha = 2 $, where $\rho_0$ is the residual resistivity and $A$ is the Fermi liquid coefficient. Because the prefactor $A$ is related to the Sommerfeld coefficient of specific heat $\gamma$ by the Kadowaki--Woods (KW) relation, $A = R_{\rm KY}\gamma^2$ with $R_{\rm KY}\sim10^{-5}$ $\mu \Omega\cdot$cm (K$^2\cdot$mol/mJ)$^2$ \cite{Kontani2008}, we can estimate the effective mass from the $A$ value for $n \geq 3$, where $T^2$-behavior is observed. As shown in Fig. \ref{13SL_QCP}(a), the prefactor $A$ becomes larger with decreasing $n$; consequently, the quasiparticle mass appears to be enhanced toward $n = 2$. Specifically, the $\gamma$ value estimated from $A$ is enhanced from the bulk CeIn$_3$ value of $\sim 120$ mJ/(K$^2$ mol) and reaches $\sim 350$ mJ/(K$^2$ mol) for $n = 3$, which corresponds to a quasiparticle mass at least several hundred times larger than the free electron mass. This mass enhancement indicates the existence of a QCP for $n<3$.

An important question remains: what is the dimensionality and type of the QCP observed in CeIn$_3$/LaIn$_3$ superlattices? The temperature exponent $\alpha$ of the resistivity changes depending on the dimensionality and type of fluctuations in the vicinity of the QCP. Now, we examine $\alpha$ for $n=2$, around which a QCP exists. As shown in Fig. \ref{13SL_QCP}(b), a $T$-linear dependence with $\alpha = 1.01\pm 0.02$ is observed for $n = 2$, whereas a $T^2$-dependence of $\rho$ is observed at low temperatures for $n \geq 3$. The $T$-linear behavior is consistent with scattering by the fluctuations enhanced by the 2D-AFM QCP \cite{Moriya2000}. This case is in contrast to the 3D case, where $\alpha$ is expected to be 1.5, and indeed $\alpha\sim 1.6$ is observed near the pressure-induced QCP in the case of bulk CeIn$_3$. 

For $n = 1$, the determination of $\alpha$ is challenging because of the weak temperature dependence of $\rho$. However, according to the literature, quantum criticality in strongly correlated electron systems can also be assessed by observing their properties under an external magnetic field \cite{Gegenwart2008,Bianchi2003,Shibauchi2008}. In fact, quantum criticality for $n = 2$ is removed by strong magnetic fields: the Fermi-liquid properties with $\alpha = 2$ are recovered, and the $A$ value is simultaneously quickly suppressed with increasing magnetic field $H$ (Fig. \ref{13SL_QCP}(a)). In the case of $n=1$, the resistivity is also strongly affected by magnetic fields: a negative magnetoresistance is observed, consistent with the view that AFM fluctuations responsible for the scattering are suppressed by the magnetic field. 

Thus, both the temperature and field dependencies of the transport properties in CeIn$_3$/LaIn$_3$ superlattices provide strong support for the presence of the quantum phase transition in the vicinity of $n=2$, which is the outcome of dimensionality tuning. Furthermore, we conclude that the reduced dimensionality enhances the fluctuations associated with the 2D AFM QCP and hence suppresses the magnetic order. The successful growth of epitaxial Ce-based superlattices promises to provide a platform for exploring the fundamental physics of strongly correlated electron systems such as the 2D Kondo lattice and, potentially, 2D superconductivity near a QCP.

\begin{figure*}[bt]
\begin{center}
\vspace{2mm}
\includegraphics[width=0.95\linewidth]{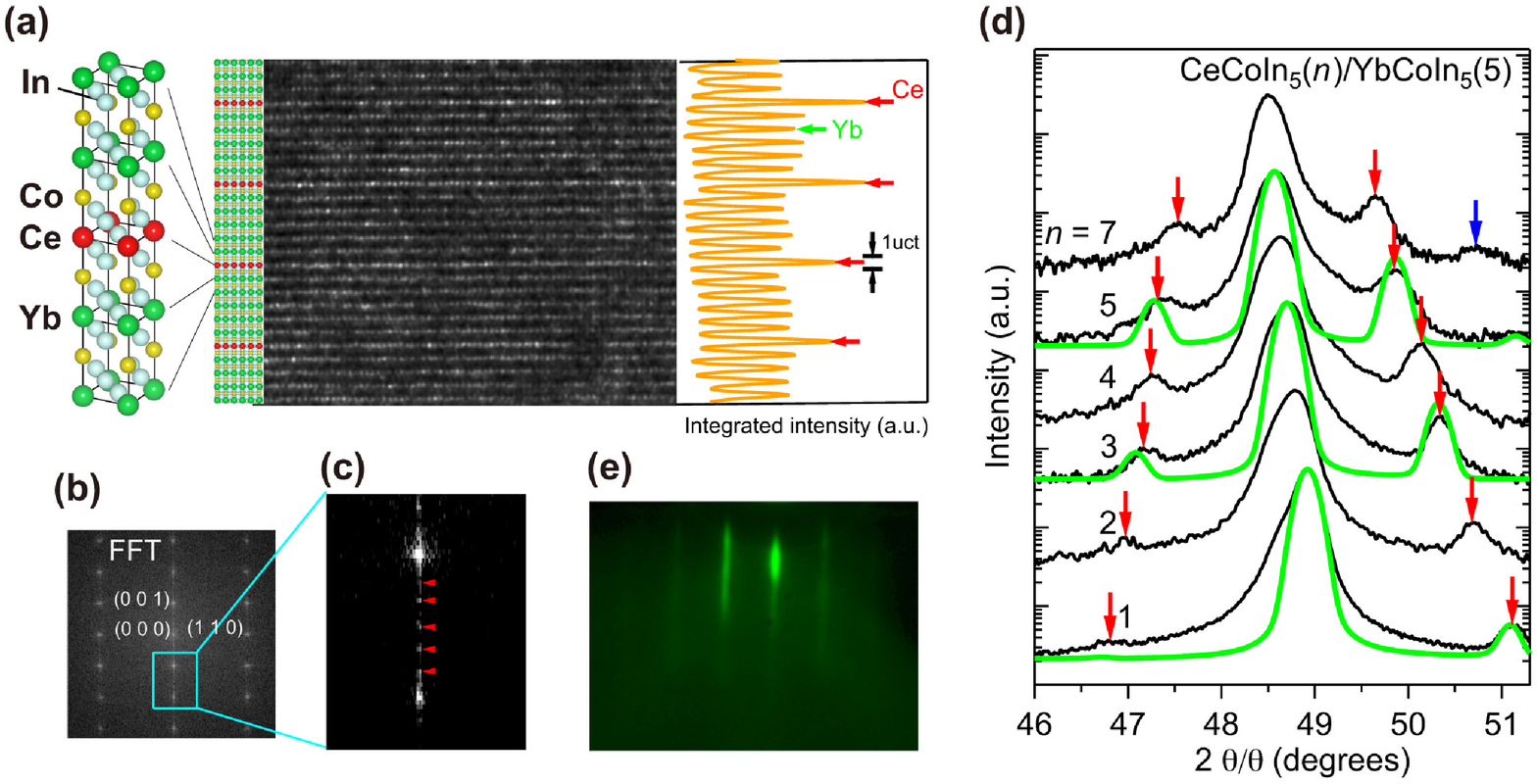}
\caption{Epitaxial superlattices ($n$:$5$) of CeCoIn$_5$($n$)/YbCoIn$_5$($5$). (a) High-resolution cross-sectional TEM image of n = 1 superlattice. The bright dot arrays are identified as the Ce layers and the less-bright dots are Yb atoms. The left panel of the TEM image represents the designed superlattice structure. The right panel represents the  intensity integrated over the horizontal width of the TEM image plotted against vertical position. (b) The fast Fourier transform (FFT) of the TEM image. (c) Magnified view shows clear superspots along the [001] direction. (d) Cu K$\alpha$1 X-ray diffraction patterns of the ($n$:$5$) superlattices for $n = 1$, $2$, $3$, $4$, $5$, and $7$ show first (red arrows) and second (blue arrow) satellite peaks. Both the positions and the asymmetric heights of the satellite peaks can be reproduced by the step-model simulations \cite{Fullerton1992} (green lines), which neglect interface and layer-thickness fluctuations. (e) Streak patterns of the RHEED image during the growth.}
\label{115SL_characterization}
\end{center}
\vspace{-5mm}
\end{figure*}

\section{2D heavy fermion superconductor of CeCoIn$_5$/YbCoIn$_5$ superlattices}
We next investigate how the superconducting state evolves with the dimensionality reduction of the Kondo lattice. Although CeCoIn$_5$ is a candidate for 2D heavy fermion superconductor, all the electronic, magnetic, and superconducting properties measured for bulk CeCoIn$_5$ are essentially 3D rather than 2D. Therefore, whether superconductivity would persist upon the dimensionality reduction of the Kondo lattice from 3D to 2D is unclear. Here, we report our observation of superconductivity in a strongly correlated system of heavy electrons confined within a 2D Kondo lattice \cite{Mizukami2011}, which was realized by fabricating superconducting superlattices built of alternating layers of $n$-UCT CeCoIn$_5$ and $m$-UCT YbCoIn$_5$, forming an ($n$:$m$) heterostructure, {\it i.e.}, An AB-type superlattice. Here, we fix $m=5$ to satisfy the condition for a 2D Kondo lattice, in which the interlayer magnetic interaction is negligibly small.

\subsection{Characterization of CeCoIn$_5$/YbCoIn$_5$ superconducting superlattices}
The high-resolution cross-sectional TEM micrographs of the ($n$:$m$)=($1$:$5$) superlattice (Fig. \ref{115SL_characterization}(a)), in which the bright spots are identified as the Ce layers and the darker spots are Yb atoms, is consistent with the designed superlattice structure shown on the left side of the TEM image. The intensity integrated over the horizontal width of the TEM image indicates a clear difference between Ce and Yb layers, showing no discernible atomic inter-diffusion between neighboring Ce and Yb layers. The fast Fourier transform (FFT) of the TEM image (Figs. \ref{115SL_characterization}(b) and (c)) also shows clear superspots along the [001] direction, which reflects the long-period stacking structure of the superlattices. Thus, the TEM results provide clear evidence for the formation of an ($n$:$m$) = ($1$:$5$) superlattice. For all superlattices with various $n$ values, the X-ray diffraction patterns (Fig. \ref{115SL_characterization}(d)) show distinct lateral satellite peaks, demonstrating that the CeCoIn$_5$ layers are continuous and evenly spaced, with no discernible inter-diffusion. The epitaxial growth of each layer with atomic flatness was confirmed by the streak pattern of the RHEED image (Fig. \ref{115SL_characterization}(e)). These results indicate the successful fabrication of epitaxial superlattices with sharp interfaces.

\begin{figure*}[bt]
\begin{center}
\vspace{2mm}
\includegraphics[width=0.95\linewidth]{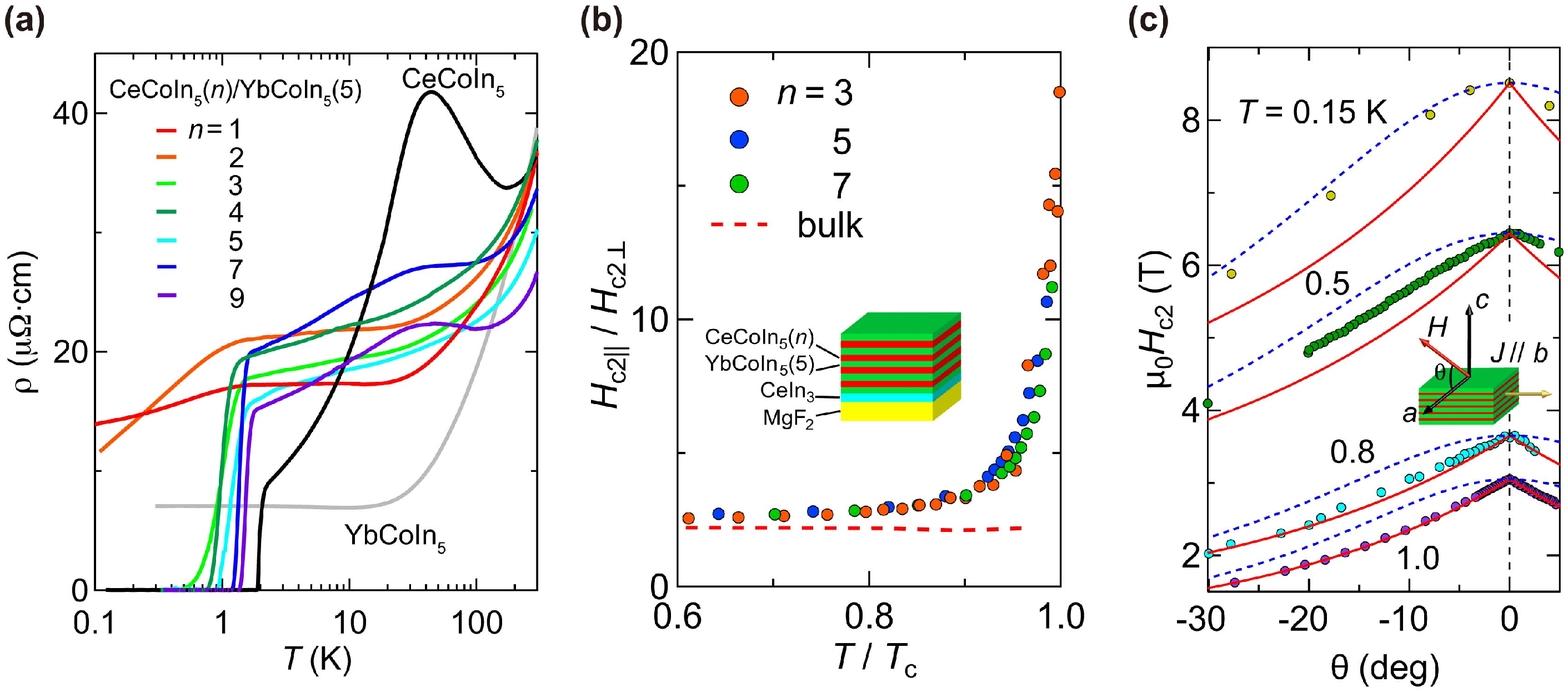}
\caption{Superconductivity in superlattices ($n$:$5$) of CeCoIn$_5$($n$)/YbCoIn$_5$($5$). (a) Temperature dependence of the resistivity $\rho(T)$ for the $n=1$, $2$, $3$, $4$, $5$, $7$, and $9$ superlattices, along with those of CeCoIn$_5$ and YbCoIn$_5$ thin films with a thickness of 300 nm. (b) Anisotropy of the upper critical field, $H_{\rm c2\parallel}/H_{\rm c2\perp}$, as a function of the reduced temperature, $T/T_c$, for the $n=3$, $5$, and $7$ superlattices and for the bulk CeCoIn$_5$. (c) Angular dependence of the upper critical field, $H_{\rm c2}({\theta})$, at several temperatures. The dashed blue and solid red lines represent the fits to the data using the anisotropic model and the Tinkham model (see Eq. (\ref{Angle_H})), respectively.}
\label{115SL_SC}
\end{center}
\vspace{-5mm}
\end{figure*}

\subsection{2D heavy fermion superconductivity}
We investigated the transport properties of the ($n$:$5$) superlattices as a function of CeCoIn$_5$ layer thickness $n$. In a CeCoIn$_5$ thin film, the resistivity $\rho(T)$ (Fig. \ref{115SL_SC}(a)) reproduces various features, as well observed in bulk single crystals; {\it i.e.}, $\rho(T)$ shows an increase because of the Kondo scattering below $\sim 100$ K, followed by a maximum at the coherent temperature $T_{\rm coh}\sim 40$ K and a sharp decrease at the superconducting transition. In the superlattices, the coherent peak is observed; however, it becomes less pronounced, particularly when $n$ is reduced, as shown in Fig. \ref{115SL_SC}(a). At lower temperatures, superconductivity with zero resistance is clearly observed in the superlattices with $n \geq 3$. By contrast, for $n=2$ and $1$, $\rho(T)$ decreases below $\sim 1$ K but does not reach zero at the lowest temperature of our measurement. However, when the magnetic field is applied perpendicular to the layers for $n = 1$, $\rho(T)$ increases and recovers to the value extrapolated above $1$ K at $5$ T, whereas the reduction of $\rho(T)$ below $1$ K remains in the parallel field of $6$ T. Such large and anisotropic field response of $\rho(T)$ is typical for layered superconductors, demonstrating superconductivity even for $n = 1$.

Important questions are whether the superconducting electrons in the superlattices are heavy and, if so, what their dimensionality is. To answer these questions, we focus on the measurements of the upper critical field ($H_{\rm c2}$) in CeCoIn$_5$/YbCoIn$_5$ superlattices. 
The orbital upper critical field at zero temperature for the field perpendicular to the layers, $H^{\rm orb}_{\rm c2\perp}(0)$, which reflects the effective electron mass in the plane, is estimated to be $6$, $11$, and $12$ T for the $n = 3$, $5$, and $7$ superlattices on the basis of the Werthamer--Helfand--Hohenberg (WHH) formula, $H^{\rm orb}_{\rm c2\perp}(0) = -0.69T_c(dH_{\rm c2\perp}/dT)_{T_c}$ \cite{Werthamer1966}. These magnitudes are comparable to or of the same order as $H^{\rm orb}_{\rm c2\perp}(0)$ ($=14$ T) estimated in the bulk CeCoIn$_5$ single crystal, providing strong evidence that heavy $f$-electrons form Cooper pairs in the superlattices. 

The superconducting order parameters of the CeCoIn$_5$ BLs are expected to be weakly coupled to each other by the proximity effect through the normal-metal YbCoIn$_5$ BLs. However, the proximity effect is negligible because the large Fermi velocity mismatch across the interface between CeCoIn$_5$ and YbCoIn$_5$ layers leads to a huge suppression of the transmission probability of electron currents \cite{She2012}. Therefore, when the thickness of the CeCoIn$_5$ BL is comparable to the perpendicular coherence length $\xi_{\perp}\sim 3$--$4$ nm for CeCoIn$_5$ (Ref.\,\onlinecite{Ikeda2001}), each CeCoIn$_5$ BL effectively acts as a 2D superconductor \cite{Schneider1993}. Indeed, the 2D feature is revealed by the diverging $H_{\rm c2 \parallel}$/$H_{\rm c2\perp}$ of the $n = 3$, $5$, and $7$ superlattices on approaching $T_c$ (Fig. \ref{115SL_SC}(b)), in sharp contrast to the result observed in the case of bulk CeCoIn$_5$ (Ref.\,\onlinecite{Tayama2002}). On the basis of these results, we conclude that 2D heavy fermion superconductivity is realized in CeCoIn$_5$($n$)/YbCoIn$_5$($5$) superlattices with $n = 3$, $5$, and $7$.

\subsection{Enhancement of $H_{\rm c2}$/$T_c$}
We now discuss the 2D superconducting properties of CeCoIn$_5$/YbCoIn$_5$ superlattices. In the 2D limit, the angular dependence of the upper critical field, $H_{c2}$($\theta$), is generally described by the Tinkham model \cite{Tinkham1963} and exhibits a cusp-like dependence near the magnetic field parallel to the plane. However, at lower temperatures, $H_{c2}$($\theta$) of the $n=3$ superlattice does not show the cusp structure and is qualitatively described by the anisotropic model \cite{Tinkham1996} (Fig. \ref{115SL_SC}(c)). This result implies that Pauli pair-breaking is dominant rather than orbital pair-breaking in the case of the parallel magnetic field. Moreover, $H_{c2\perp}(T)$ extrapolated to $T=0$ K is significantly smaller than the estimated orbital-limited upper critical field $H^{\rm orb}_{c2\perp}(0)$, indicating a predominant Pauli paramagnetic pair-breaking effect even in the case of the perpendicular field. Thus, 2D heavy fermion superconductivity of CeCoIn$_5$/YbCoIn$_5$ superlattices is also dominated by Pauli paramagnetism, similar to the case of bulk CeCoIn$_5$.

\begin{figure}[bt]
\begin{center}
\vspace{2mm}
\includegraphics[width=0.95\linewidth]{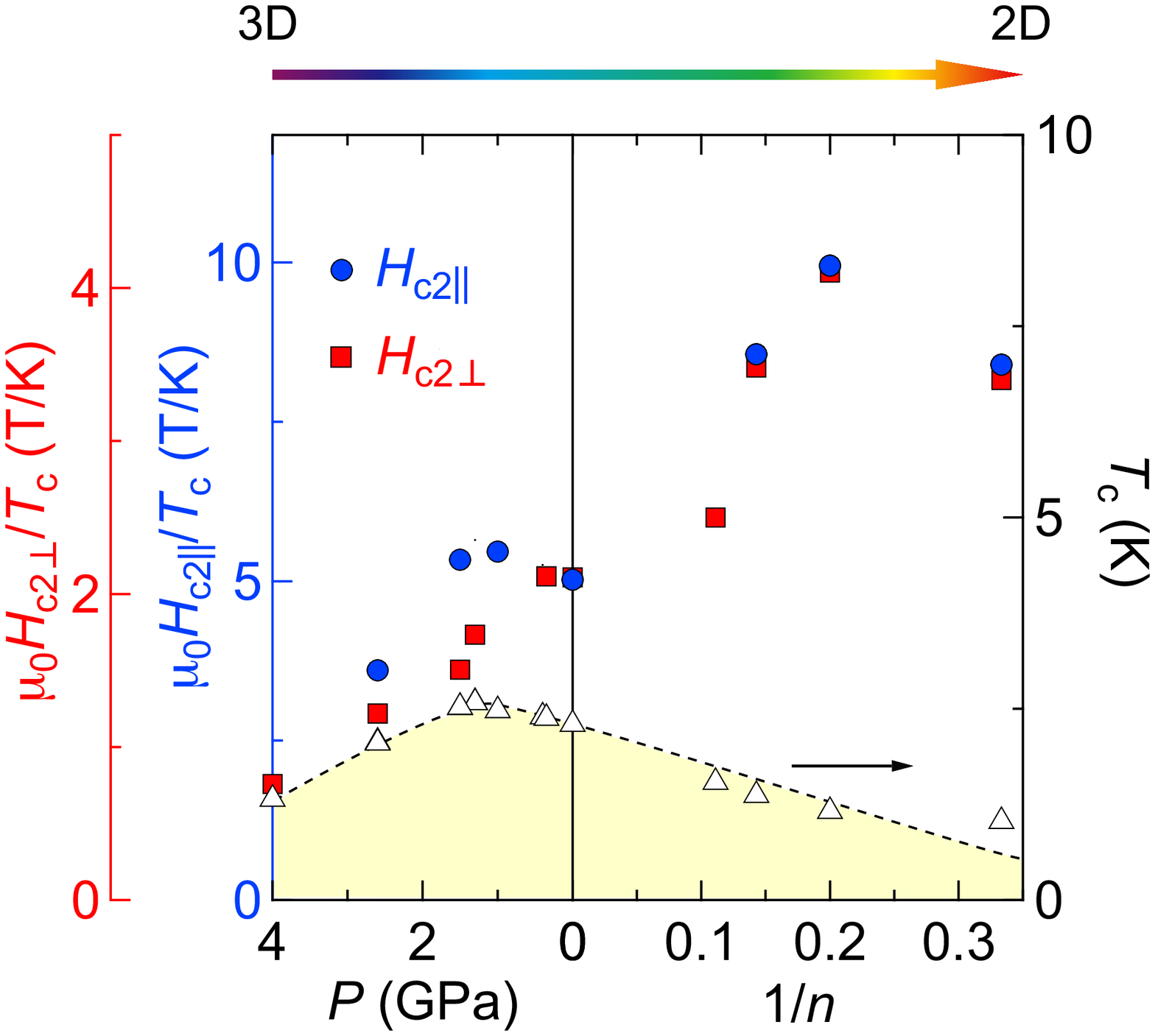}
\caption{ Superconducting transition temperature, $T_c$ (open triangles), and the upper critical fields divided by the transition temperature, $H_{\rm c2}$/$T_c$, in parallel (filled blue circles) and perpendicular (filled red squares) fields as a function of dimensionality parameter 1/$n$ (right panel). The pressure dependence of these quantities is also shown for comparison (left panel). The dimensionality is expected to increase with the pressure. Note the different scales for the parallel (blue) and the perpendicular (red) fields.}
\label{115SL_Hc2_Tc}
\end{center}
\vspace{-5mm}
\end{figure}

The most remarkable feature of 
this 2D heavy fermion superconductor is that $T_c$ decreases rapidly from the bulk value with decreasing $n$, whereas $H_{c2}$ values hardly change for either field direction at low temperatures. Indeed, $H_{c2}$ of $n = 5$ and $7$ is comparable to or greater than that of the bulk at low temperatures. This result immediately implies that the thickness reduction drastically enhances $H_{c2}/T_c$ from the bulk value, as shown in Fig. \ref{115SL_Hc2_Tc}. The pressure-dependence results \cite{Knebel2010}, which are expected to show CeCoIn$_5$ as more three-dimensional, also exhibit a general trend of enhanced $H_{c2}/T_c$ with reduced dimensionality, although $T_c$ exhibits a dome-shaped dependence. 
Until now, three hypotheses have been put forward to explain this trend: (a) extremely strong-coupling superconductivity, (b) an FFLO state, and (c) locally noncentrosymmetric superconductivity. In the following, we briefly introduce the details of these scenarios.

\vspace{5mm}
\begin{center}
{\noindent}{(a) Extremely strong-coupling superconductivity}
\end{center}

In CeCoIn$_5$/YbCoIn$_5$ superlattices, $H_{c2}$ is expected to be described by the Pauli-limited upper critical field $H^{\rm P}_{c2}$, which is given by
\begin{equation}
H^{\rm P}_{c2}=\sqrt{2}\Delta/g\mu_B \cite{Chandrasekhar1962,Clogston1962},
\label{H_Pauli}
\end{equation}
where $g$ is the gyromagnetic ratio determined by the Ce crystalline electric field levels. The $g$-value of the superlattices is assumed to deviate little from the bulk value because the anisotropy of $H_{c2}$ at low temperatures does not depend on $n$, as shown in Fig. \ref{115SL_SC}(b). In this case, via the relation given in Eq. (\ref{H_Pauli}), the striking enhancement of $H_{c2}/T_c$ with decreasing $n$ implies a remarkable enhancement of $\Delta/T_c$ by two-dimensionalization. Using the reported value of $2\Delta/k_BT_c = 6$ in the bulk single crystal \cite{WPark2008}, we estimate $2\Delta/k_BT_c$ for the $n = 5$ superlattice to exceed $10$, which is significantly enhanced from  the weak-coupling Bardeen--Cooper--Schrieffer value of $2\Delta/k_BT_c = 3.54$. This result is also supported by the theory that $d$-wave pairing mediated by AFM fluctuations in 2D can be much stronger than that in 3D \cite{Takimoto2002,Arita1999,Monthoux2001}. Thus, extremely strong-coupling superconductivity may be realized in the 2D Kondo lattice.

\vspace{5mm}
\begin{center}
{\noindent}{(b) Fulde--Ferrell--Larkin--Ovchinnikov (FFLO) state}
\end{center}

In the normal state, an external magnetic field splits the spin-up and spin-down states (Zeeman splitting); hence, the free energy of the system can be reduced by polarizing the spin of the electrons. By contrast, spin-singlet Cooper pairs are not spin polarized by the Zeeman effect because the pairs are formed between spin-up and spin-down electrons. Therefore, to polarize condensed electrons, Cooper pairs usually must be broken within the framework of the BCS theory, where the total momentum of the pairs is zero. However, if the formation of a new pairing state (${\mbox{\boldmath $k$}}\uparrow$, ${-\mbox{\boldmath $k$}}+{\mbox{\boldmath $q$}}\downarrow$) with finite ${\mbox{\boldmath $q$}}$ is realized, the system can lower its free energy even in the presence of a magnetic field. This state is called the FFLO state \cite{Fulde1964,Larkin1965}, which suppresses the Pauli pair-breaking effect. 

The FFLO state is suggested to be more stable when a large area of the spin-up Fermi surface is connected to the spin-down surface by the ${\mbox{\boldmath $q$}}$-vector \cite{Matsuda2007}. This effect bears a striking resemblance to the nesting effect of charge-density-wave (CDW) and spin-density-wave (SDW) transitions. For a 3D system with a spherical Fermi surface, the spin-down and spin-up Fermi surfaces touch only at a point by a shift of the ${\mbox{\boldmath $q$}}$-vector. For a 2D system with a cylindrical Fermi surface, the two Fermi surfaces touch on a line by a shift of the ${\mbox{\boldmath $q$}}$-vector. Therefore, two-dimensionality generally favors the FFLO state, leading to the suppression of Pauli paramagnetic pair-breaking.

\vspace{5mm}
\begin{center}
{\noindent}{(c) Locally noncentrosymmetric superconductivity}
\end{center}

In the absence of space inversion symmetry, an asymmetric potential gradient $\nabla V$ yields a spin-orbit interaction \cite{Bauer2012}. For instance, asymmetry of the potential in the direction perpendicular to the 2D plane $\nabla V \parallel [001]$ induces Rashba spin-orbit interaction \cite{Rashba1960}, $\alpha_R {\bm g}({\bm k})\cdot{\bm \sigma}\propto({\bm k} \times \nabla V)\cdot {\bm \sigma}$, where ${\bm g}({\bm k})=(-k_y,k_x,0)/k_F$, $k_F$ is the Fermi wave number, and ${\bm \sigma}$ is the Pauli matrix. The Rashba interaction splits the Fermi surface into two sheets with different spin structures: the spin direction is tilted into the plane, rotating clockwise on one sheet and anticlockwise on the other. In this case, the Zeeman interaction for a magnetic field can be well approximated by $\pm {\bm g}({\bm k})\cdot \mu_0{\bm H}$, leading to a strong suppression of the Pauli effect, in particular for ${\bm H}\parallel c $ where ${\bm g}({\bm k})$ is always perpendicular to ${\bm H}$ \cite{Gor'kov2001,Frigeri2004,Mineev2011}.

Recently, local ISB at the interfaces of multilayer systems has been proposed to also suppress the Pauli pair-breaking effect when the interlayer hopping integral is similar to or smaller than the Rashba splitting ($t_c \alt \alpha_R$) \cite{Maruyama2012}. The reason is that, because their coupling is expected to be weak because of the violation of spin and momentum conservation, noncentrosymmetric interface layers and centrosymmetric inner layers can be treated independently. In the CeCoIn$_5$/YbCoIn$_5$ superlattice, the inversion symmetry is {\it locally} broken at the top and the bottom ``interface" CeCoIn$_5$ layers in the immediate proximity of the YbCoIn$_5$ BLs \cite{Goh2012}, as illustrated in Fig. \ref{Kondo_SL}(a). Although the inner layers might also break the local inversion symmetry, the effect has been theoretically demonstrated to be much weaker. Thus, with the reduction of $n$, the fraction of noncentrosymmetric interface layers increases rapidly. In the presence of the local ISB, coupled with the fact that Ce has a large atomic number, Rashba-type spin-orbit coupling is expected to be strong, which appears to satisfy the condition $t_c \alt \alpha_R$. Therefore, the enhancement of spin susceptibility in the superconducting state has been predicted with decreasing $n$, leading to the suppression of the Pauli pair-breaking effect. At present, this scenario is the prevailing one for the enhancement of $H_{c2}/T_c$, as discussed in the next section.

\subsection{Local inversion symmetry breaking}
Recently, the importance of local ISB for the superconducting state has been emphasized experimentally through the anomalous temperature and angular dependencies of $H_{c2}$ \cite{Goh2012,Shimozawa2014}, which can be interpreted as a strong suppression of the Pauli pair-breaking effect. Recent site-selective NMR spectroscopy experiments have also indicated that the local ISB affects the magnetic properties, leading to the suppression of AFM fluctuations \cite{Yamanaka2015}. Thus, the interfacial effect becomes a key element for understanding the electronic state in the CeCoIn$_5$/YbCoIn$_5$ superlattices. The details of these experiments are given in the following subsections.

\subsubsection{Superconducting properties}
\begin{figure}[tb]
\begin{center}
\vspace{2mm}
\includegraphics[width=0.85\linewidth]{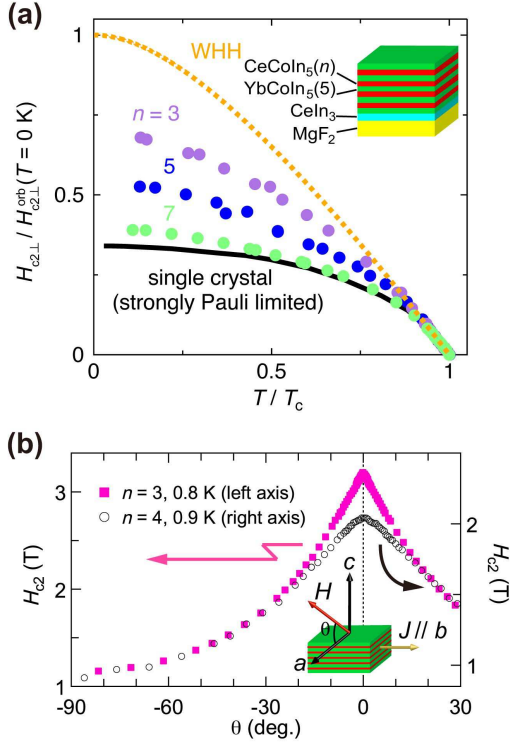}
\caption{Upper critical fields in the CeCoIn$_5$($n$)/YbCoIn$_5$($5$) superlattices. (a) $H_{c2\perp}/H_{c2\perp}^{orb}(0)$ as a function of $T/T_c$ for the $n=3$, $5$, and $7$ superlattices. Also plotted are $H_{c2\perp}/H_{c2\perp}^{orb}(0)$ for a CeCoIn$_5$ single crystal with a strong Pauli effect and the WHH curve without the Pauli effect. (b) Comparison of $H_{c2}(\theta)$ for the $n=3$ (pink filled squares, left axis) and $n=4$ (black open circles, right axis) superlattices near $T_c$.}
\label{Hc2_local}
\end{center}
\vspace{-5mm}
\end{figure}

Figure \ref{Hc2_local}(a) shows the temperature dependence of the perpendicular upper critical field, $H_{c2\perp}$, normalized by $H_{c2\perp}^{\rm orb}(0)$ for the superlattices CeCoIn$_5$($n$)/YbCoIn$_5$(5) with $n$ = 3, 5, and 7. We also include two extreme cases: $H_{c2\perp}/H_{c2\perp}^{\rm orb}(0)$ for the bulk CeCoIn$_5$, in which superconductivity is dominated by Pauli paramagnetism \cite{Tayama2002}, and the WHH curve without the Pauli effect \cite{Werthamer1966}. A remarkable feature is that $H_{c2\perp}/H_{c2\perp}^{\rm orb}(0)$ is enhanced with decreasing $n$ over a wide temperature range. This effect is unlikely to be attributable to impurity scattering because Yb substitution for Ce sites in the bulk Ce$_{1-x}$Yb$_x$CoIn$_5$ system does not change the overall temperature dependence of $H_{c2\perp}/H_{c2\perp}^{\rm orb}(0)$. We can also exclude the possibility that this enhancement is related to the dimensionality of the superconductivity, because $H_{c2\perp}/H_{c2\perp}^{\rm orb}(0)$ varies with $n$ even at $T\sim 0.8T_c$, where $\xi_{\perp}(T)$ is substantially longer than the Ce-BL thickness for all superlattices. The enhancement is rather associated with the reduction of the Maki parameter $\alpha_M$, which represents the ratio of the orbital-limited upper critical field to that of the Pauli-limited case, although spin-orbit scattering may change slightly with $n$. Thus, the enhancement of $H_{c2\perp}/H_{c2\perp}^{\rm orb}(0)$ with decreasing $n$ indicates that the local ISB, whose degree is enhanced with decreasing $n$, modifies the superconducting properties, leading to the relative suppression of the Pauli effect with respect to the orbital effect.

The suppression of the Pauli pair-breaking effect can be identified clearly by the angular dependence of $H_{c2}(\theta)$  (Fig. \ref{Hc2_local}(b)), where $\theta$ is the angle between $\bm{H}$ and the $a$-axis. For $n=3$, the distinct cusp behavior is observed at $\theta=0$, whereas for $n=4$ the cusp structure is not observed for all $\theta$. This remarkable difference of $H_{c2}(\theta)$ between the $n=3$ and $n=4$ superlattices is highly unusual because the CeCoIn$_5$ BLs in each superlattice have similar thicknesses; hence, similar angular variation of $H_{c2}(\theta)$ is expected, in particular near $T_c$ where $\xi_{\perp}(T)$ well exceeds the CeCoIn$_5$ BL thickness. To describe the difference quantitatively, we analyzed the data using the model \cite{Goh2012}:
\begin{equation}
\left[\frac{H_{c2}(\theta)\cos\theta}{H_{c2}(0^{\circ})}\right]^2+\beta_{T}\left|\frac{H_{c2}(\theta)\sin\theta}{H_{c2}(90^{\circ})}\right|+\beta_{P}\left[\frac{H_{c2}(\theta)\sin\theta}{H_{c2}(90^{\circ})}\right]^2=1,
\label{Angle_H}
\end{equation}
where $\beta_{T}$ ($\geq$ 0) and $\beta_{P}$ ($\geq$ 0) are fitting parameters with $\beta_{T}+\beta_{P}=1$; $(\beta_{T},\beta_{P})=(1,0)$ represents the so-called Tinkham model \cite{Tinkham1996}, which describes $H_{c2}(\theta)$ in a 2D thin film in the absence of the Pauli effect. In the Tinkham model, a cusp appears at $\theta=0$ as a result of the vortex formation due to the orbital pair-breaking effect in a slightly tilted field, which strongly suppresses $H_{c2}$. By contrast, $(\beta_{T},\beta_{P})=(0,1)$ represents the anisotropic model, which describes $H_{c2}(\theta)$ of 2D superconductors with a strong Pauli effect. In this case, a cusp does not appear because $H_{c2}(\theta)$ given by the relation in Eq. (\ref{H_Pauli}) is determined by the anisotropy of the $g$-factor, which changes smoothly with $\theta$. Therefore, $\beta_{T}/\beta_{P}$ quantifies the relative contribution of orbital and Pauli paramagnetic effects. $H_{c2}(\theta)$ is well fitted with this model,  and $\beta_{T}/\beta_{P}$ is obtained. As depicted in Fig. \ref{Hc2_block layer}(c), $\beta_{T}/\beta_{P}$ is strongly enhanced with decreasing $n$, indicating the suppression of the Pauli effect. This result is consistent with the enhancement of $H_{c2\perp}/H_{c2\perp}^{\rm orb}(0)$ with decreasing $n$ shown in Fig. \ref{Hc2_local}(a). Thus, both the enhancement of $H_{c2\perp}(T)/H_{c2\perp}^{\rm orb}(T=0)$ in a perpendicular field and the angular variation of $H_{c2}(\theta)$ around a parallel field indicate that the local ISB strongly affects the superconductivity through the suppression of Pauli paramagnetism.

\subsubsection{Magnetic properties}
\begin{figure}[tb]
\begin{center}
\vspace{2mm}
\includegraphics[width=0.85\linewidth]{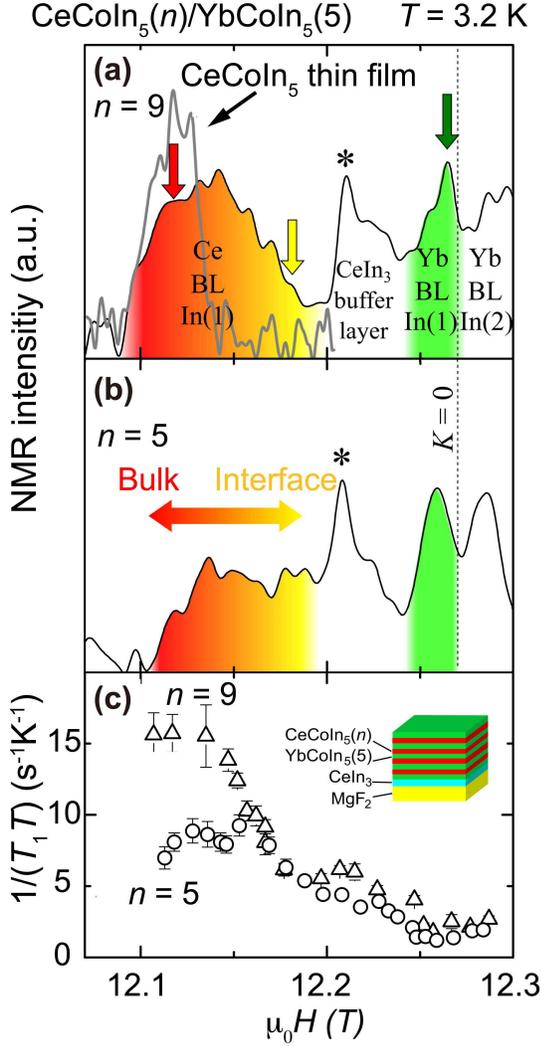}
\caption{Site-selective NMR measurements of the superlattices. (a), (b) $^{115}$In(1)-NMR spectra of the $n=9$ and $5$ superlattices at $T=3.2$ K and with fixed NMR frequency ($f=114.5$ MHz). The red-to-yellow gradation represents the NMR signal from inner to interface layers in CeCoIn$_5$ BLs. (c) Field dependence of ($T_1T$)$^{-1}$ in the $n = 9$ and $5$ superlattices.}
\label{NMR}
\end{center}
\vspace{-5mm}
\end{figure}

Figures \ref{NMR}(a) and \ref{NMR}(b) show the NMR spectra at 3.2 K near the In(1) central transition for the $n = 9$ and $n = 5$ superlattices. We can separate signals from three different layers by comparing the spectra with those of CeCoIn$_5$ and YbCoIn$_5$ thin films: the red, orange, and yellow shaded regions represent the spectra of CeCoIn$_5$ BLs, the green regions the spectra of YbCoIn$_5$ BLs, and the peak shown by an asterisk the spectra of CeIn$_3$ buffer layers. A notable feature is that, in the lower-field region, the spectrum of CeCoIn$_5$ BLs for $n=9$ has a much larger weight compared to that for $n = 5$. This observation implies that the spectral regions shaded by yellow and red arise from the interface and from the inner layers of CeCoIn$_5$ BLs, respectively, because the volume fraction of the interface layers increases rapidly with decreasing $n$. Therefore, the field dependence of $(T_1T)^{-1}$ allows the layer dependence of the AFM fluctuations to be resolved even within the same CeCoIn$_5$ BL because $(T_1T)^{-1}$ at the In(1) detects the magnitude of the fluctuations. 

The field dependence of $(T_1T)^{-1}$ (Fig. \ref{NMR}(c)) indicates that AFM fluctuations near the interface are weaker than those at the inner layers of CeCoIn$_5$ BLs. This strong spatial dependence implies that the interfacial effect plays a critical role in determining the magnetic properties of the CeCoIn$_5$ BLs. The proximity of $f$-electrons with a magnetic moment to nonmagnetic YbCoIn$_5$ layers is unlikely to be the origin of the reduction of the fluctuations near interfaces because magnetic fluctuations in the YbCoIn$_5$ BLs are essentially the same as those in the YbCoIn$_5$ thin film. Therefore, the magnetic degrees of freedom of $f$-electrons are quenched inside the YbCoIn$_5$ BLs. A possible explanation for the weakening of the AFM fluctuations is the effect of the local ISB at the interfaces between CeCoIn$_5$ and YbCoIn$_5$ BLs. In this case, the Fermi surface splitting due to the Rashba spin-orbit interaction should seriously change the nesting condition; hence, the commensurate AFM fluctuations with ${\bm Q} = (1/2,1/2,1/2)$, which is dominant in bulk CeCoIn$_5$ \cite{Stock2008}, is expected to be suppressed. Moreover, as noted in the literature, the local ISB reduces the AFM fluctuations by lifting the degeneracy of the fluctuation modes through the helical anisotropy of the spin configuration shown in Fig. \ref{Kondo_SL}(a) \cite{Yanase2008,Takimoto2008}. Therefore, the reduction of $(T_1T)^{-1}$ near interfaces implies the significance of the local ISB for the magnetic properties of CeCoIn$_5$/YbCoIn$_5$ superlattices.

\begin{figure*}[t]
\begin{center}
\vspace{2mm}
\includegraphics[width=0.8\linewidth]{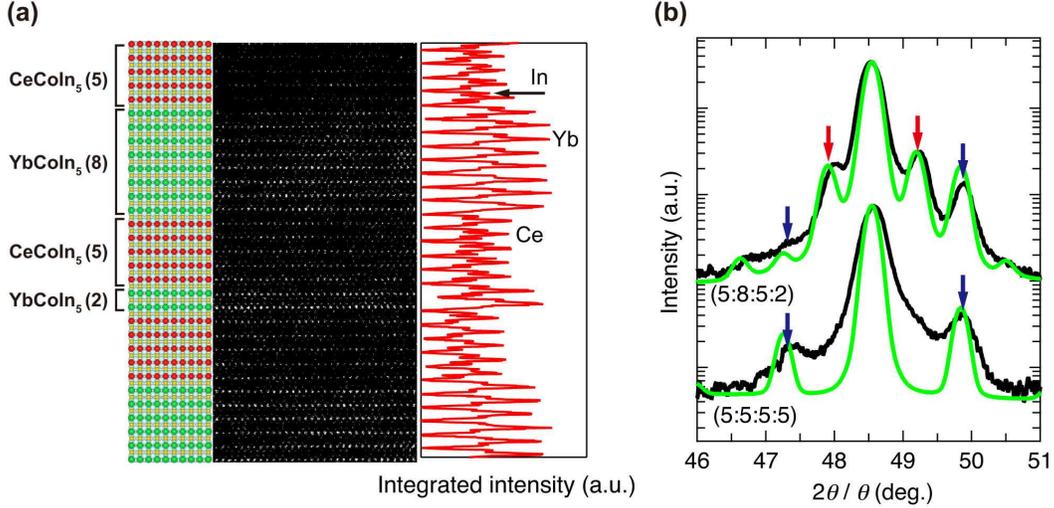}
\vspace{-5mm}
\caption{Modulated superlattices ($5$:$m$:$5$:$m^{\prime}$) of CeCoIn$_5$($5$)/YbCoIn$_5$($m$)/CeCoIn$_5$($5$)/YbCoIn$_5$($m^{\prime}$). (a) HAADF-STEM image of the ($n$:$m$:$n$:$m'$) = (5:8:5:2) superlattice. The bright spots correspond to Yb atoms, and the faint spots are Ce atoms. The left image represents the designed (5:8:5:2) superlattice structure. The right panel represents the intensity integrated over the horizontal width of the HAADF-STEM image. (b) Cu K$\alpha$1 X-ray diffraction patterns in the 2$\theta$ region around the (004) peak for ($n$:$m$:$n$:$m'$) = (5:5:5:5) and (5:8:5:2) superlattices with a typical thickness of 300 nm. Two types of satellite peaks (marked by blue and red arrows) appear. Green lines represent the step model simulations in which interface and layer-thickness fluctuations are ignored.}
\label{modulation_characterization}
\end{center}
\vspace{-5mm}
\end{figure*}

\vspace{5mm}
To summarize this section, we have clarified that the CeCoIn$_5$/YbCoIn$_5$ superlattice shows Pauli-limited superconductivity, where heavy $f$-electrons confined within a 2D Kondo lattice form Cooper pairs. Moreover, $H_{c2\perp}(T)$, $H_{c2}(\theta)$, and $(T_1T)^{-1}$ results revealed that the local ISB at interfaces is a key factor for understanding both superconducting and magnetic properties in the CeCoIn$_5$/YbCoIn$_5$ superlattice, even though the global inversion symmetry is preserved in the whole crystals. This idea is also supported by very recent experiments of AFM Kondo superlattices CeRhIn$_5$/YbRhIn$_5$ \cite{Ishii2016}, where the suppression of the Zeeman effect by local ISB helps reduce the destruction of the AFM state under the magnetic field. Thus, the successful fabrication of CeCoIn$_5$/YbCoIn$_5$ superlattice appears to be the first step toward understanding the combination of $f$-electron physics, superconductivity, and low dimensionality as well as interfacial effects. In fact, these complexities are expected to give rise to novel superconducting states in heavy fermion systems, such as the interplay of Kondo lattice physics and the Berezinskii--Kosterlitz--Thouless (BKT) transition \cite{She2012}, dimensional-crossover in a heavy fermion superconductor \cite{Tada2013,Tada2015}, 2D FFLO phases \cite{Matsuda2007}, and superconductivity accompanied by symmetry breaking not available in the bulk such as the local ISB \cite{Maruyama2012}.

\section{Controllable Rashba spin-orbit interaction in CeCoIn$_5$/YbCoIn$_5$ superlattices}
To examine more widely how ISB affects the superconductivity of the 2D CeCoIn$_5$ BLs, we proceeded one step further by fabricating {\it modulated} superlattices \cite{Shimozawa2014}; in this case, the thickness of CeCoIn$_5$ is now kept to $n=5$ for the entire superlattice, whereas the thickness of YbCoIn$_5$ alternates between $m$ and $m^{\prime}$ from one block layer to the next, forming an ($5$:$m$:$5$:$m^{\prime}$) $c$-axis-oriented superlattice structure. Here, we study superlattices both without and with thickness modulation of YbCoIn$_5$ layers, $m=m^{\prime}$ (AB-type, Fig. \ref{Kondo_SL}(a)) and $m \neq m^{\prime}$ (ABAB$^{\prime}$-type, Fig. \ref{Kondo_SL}(b)), whose total number is fixed at $m+m^{\prime}=10$. In the former case, inversion symmetry is locally broken at the interfaces between CeCoIn$_5$ and YbCoIn$_5$ layers, as described in the previous section, whereas in the latter case, an additional ISB, {\it i.e.}, a ``block-layer" ISB, is introduced into the CeCoIn$_5$ BLs, in addition to the local ISB (see also the Sec. 3.C ``Superconducting Kondo superlattices").

\subsection{Characterization of the modulated superlattices}
Figure \ref{modulation_characterization}(a) shows the high-angle annular dark-field scanning transmission electron microscopy (HAADF-STEM) results for the ($n$:$m$:$n$:$m'$) = ($5$:$8$:$5$:$2$) superlattice. The bright spots correspond to Yb atoms, and the faint spots are Ce atoms, which are consistent with the designed superlattice structure shown on the left side of the HAADF-STEM image. The intensity integrated over the horizontal width of the HAADF-STEM image shown on the right side of the HAADF-STEM image indicates a clear difference between Ce and Yb layers. Thus, the HAADF-STEM results demonstrate the formation of a modulated superlattice--a ($5$:$8$:$5$:$2$) superlattice--with no discernible interdiffusion. 

The difference between the ($5$:$5$:$5$:$5$) and ($5$:$8$:$5$:$2$) superlattices is clearly observed in the X-ray diffraction pattern (Fig. \ref{modulation_characterization}(b)): the satellite peaks (marked by blue arrows) are observed in the ($5$:$5$:$5$:$5$) superlattice, which reflects the long-period structures for an AB-type superlattice, whereas extra satellite peaks (marked by red arrows) appear because of the thickness modulation of YbCoIn$_5$ in the ($5$:$8$:$5$:$2$) superlattice. The position of the satellite peaks and their asymmetric heights can be reproduced by the step model simulation \cite{Fullerton1992} (green lines) which does not include interdiffusion, confirming the fabrication of both the ($5$:$5$:$5$:$5$) and ($5$:$8$:$5$:$2$) superlattices with sharp interfaces.

\begin{figure*}[tb]
\begin{center}
\vspace{2mm}
\includegraphics[width=\linewidth]{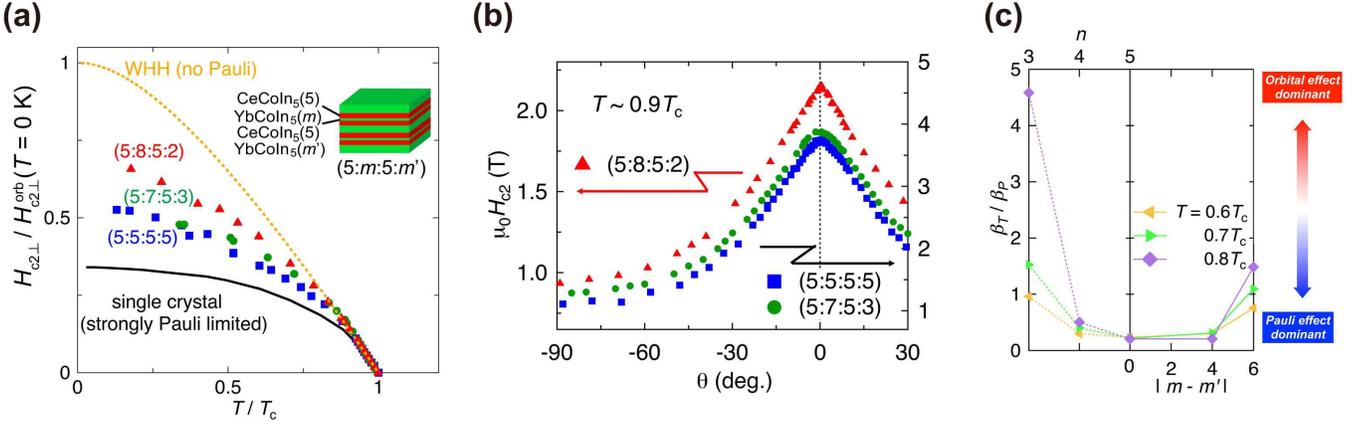}
\caption{Upper critical fields in the ABAB$^{\prime}$ superlattices. (a) $H_{c2\perp}/H_{c2\perp}^{orb}(0)$ versus $T/T_c$ for ($n$:$m$:$n$:$m'$) = (5:5:5:5), (5:7:5:3), and (5:8:5:2) superlattices, along with that for a CeCoIn$_5$ single crystal with strong Pauli effect and the WHH curve without the Pauli effect. (b) Comparison of $H_{c2}(\theta)$ in (5:8:5:2) (red triangles, left axis), (5:7:5:3) (green circles, right axis) and  (5:5:5:5) (blue squares, right axis) superlattices at around $T=0.9T_c$. (c) $\beta_{T}/\beta_{P}$, which quantifies the relative importance of the orbital and Pauli-paramagnetic pair-breaking effects, is plotted as a function of the thickness modulation of YbCoIn$_5$ layers $|m-m^{\prime}|$ (right panel). For comparison, $\beta_{T}/\beta_{P}$ in the AB-type superlattice CeCoIn$_5$($n$)/YbCoIn$_5$(5) is plotted as a function of $n$ (left panel).}
\label{Hc2_block layer}
\end{center}
\vspace{-5mm}
\end{figure*}

\subsection{2D superconducting state}
We compare the transport properties of the ABAB$^{\prime}$-type superlattices with ($n$:$m$:$n$:$m'$) = (5:7:5:3) and (5:8:5:2) with those of the AB-type superlattice with ($n$:$m$:$n$:$m'$) = (5:5:5:5). The resistivities of the AB- and ABAB$^{\prime}$-type superlattices exhibit very similar temperature dependences. The residual resistivity $\rho_0\simeq 16~\mu\Omega$ cm is independent of the thickness modulation of the YbCoIn$_5$ layers. These results imply that ISB does not affect the normal transport properties of the CeCoIn$_5$/YbCoIn$_5$ superlattices with a CeCoIn$_5$ thickness of $5$ unit-cell layers. At low temperatures, all superlattices show sharp resistive transitions to superconductivity; their transition temperatures $T_c$ are similar to one another: 1.16, 1.18, and 1.13 K for ($n$:$m$:$n$:$m'$) = (5:5:5:5), (5:7:5:3), and (5:8:5:2), respectively. Moreover, the 2D superconducting feature is revealed by the diverging $H_{c2\parallel}/H_{c2\perp}$ of the superlattices upon approaching $T_c$, in sharp contrast to the CeCoIn$_5$ thin film with a thickness of 120 nm \cite{Shimozawa2014}. Thus, these experimental results again suggest that 2D superconducting states are realized in CeCoIn$_5$/YbCoIn$_5$ superlattices, irrespective of whether inversion symmetry is present.

\subsection{Block layer inversion symmetry breaking}
We discuss the effect of Rashba splitting arising from block-layer ISB on the 2D superconductivity. Figure \ref{Hc2_block layer}(a) shows the temperature dependence of $H_{c2\perp}/H_{c2\perp}^{\rm orb}(0)$ for ($n$:$m$:$n$:$m^{\prime}$) = ($5$:$5$:$5$:$5$), ($5$:$7$:$5$:$3$) and ($5$:$8$:$5$:$2$), along with $H_{c2\perp}/H_{c2\perp}^{\rm orb}(0)$ for the bulk CeCoIn$_5$ with a strong Pauli effect \cite{Tayama2002} and the WHH curve without the Pauli effect \cite{Werthamer1966}. Remarkably, $H_{c2\perp}/H_{c2\perp}^{\rm orb}(0)$ is enhanced with increasing $|m-m^{\prime}|$ at low temperatures. This enhancement is not due to the difference in the electron scattering rate or impurity concentration because the resistivities of these superlattices exhibit very similar temperature dependences with similar $T_c$. The enhancement is rather associated with the reduction in the ratio of the orbital- to Pauli-limited upper critical fields. Thus, the present results suggest that block-layer ISB changes the superconducting properties, leading to the relative suppression of the Pauli effect with respect to the orbital effect.

To further investigate the role of the block-layer ISB, we measured the angular dependence of the upper critical field, $H_{c2}(\theta)$, where $\theta$ is the angle between $\bm{H}$ and the $a$-axis. A clear difference of $H_{c2}(\theta)$ between AB-type (5:5:5:5) and ABAB$^{\prime}$-type (5:8:5:2) superlattices is observed, as shown in Fig. \ref{Hc2_block layer}(b): for the ($5$:$8$:$5$:$2$) superlattice, $H_{c2}(\theta)$ increases monotonically without saturation as $|\theta|$ decreases, and the distinct cusp behavior is observed at $\theta=0$; by contrast, for the ($5$:$7$:$5$:$3$) and ($5$:$5$:$5$:$5$) superlattices, the cusp structure is not observed and $H_{c2}(\theta)$ is smooth for all $\theta$. To provide a more quantitative description, we analyzed the data using the model given in Eq.(\ref{Angle_H}). Figure \ref{Hc2_block layer}(c) summarizes the $\beta_{T}/\beta_{P}$ obtained by this model for several superlattices at fixed reduced temperatures. For the ABAB$^{\prime}$-type superlattices, when going from ($5$:$7$:$5$:$3$) to ($5$:$8$:$5$:$2$), $\beta_{T}/\beta_{P}$ is strongly enhanced, indicating the suppression of the Pauli effect. This result is consistent with the low-temperature enhancement of $H_{c2\perp}/H_{c2\perp}^{\rm orb}(0)$ with increasing $|m-m^{\prime}|$ shown in Fig. \ref{Hc2_block layer}(a). Thus, both the enhancement of $H_{c2\perp}(T)/H_{c2\perp}^{\rm orb}(T=0)$ in the perpendicular magnetic field and the angular variation of $H_{c2}(\theta)$ around the parallel field indicate that ISB arising from the thickness modulation of YbCoIn$_5$ layers strongly affects the superconductivity through the suppression of Pauli paramagnetism when $|m-m^{\prime}|$ is increased from $4$ to $6$. This result can be understood if Rashba splitting begins to exceed the superconducting gap energy when $|m-m^{\prime}|$ reaches a threshold value between $4$ and $6$.

Rashba splitting arising from the ISB has been suggested to affect the superconductivity, leading to several novel phenomena such as anomalous magnetoelectric effects \cite{Lu2008} and topological superconducting states \cite{Tanaka2009,Sato2009,Qi2011}. Such phenomena are expected to be more pronounced in heavy fermion systems because of a strong electron correlation effect \cite{Fujimoto2007}. Bulk CeCoIn$_5$ is one of the most attractive superconductors among these systems because it exhibits many fascinating superconducting properties, as described in Sec. 2.B ``Heavy fermion superconductor CeCoIn$_5$". Therefore, our CeCoIn$_5$-based superlattices, in which the degree of ISB and, consequently, the Rashba splitting are controllable, offer the prospect of achieving even more fascinating pairing states than the bulk CeCoIn$_5$. In fact, the availability of such superlattices has been predicted theoretically to provide an ideal playground for exploring exotic superconducting states, such as a helical vortex state \cite{Kaur2005}, pair-density-wave state \cite{Yoshida2012}, complex-stripe phases \cite{Yoshida2013}, parity-mixed superconductivity \cite{Yoshida2014}, a topological crystalline superconducting state protected by mirror symmetry \cite{Yoshida2015,Watanabe2015}, and Majorana fermion excitations \cite{Sato2010} in strongly correlated electron systems. Recently, the formation of a Majorana flat band in the present ABAB$^{\prime}$-type superlattice structure has also been proposed \cite{Yuan2014}.

\section{Conclusion}
In this report, we reviewed the research into the 2D Kondo lattice systems, which can be realized by fabricating Kondo superlattices. In the first Kondo superlattice, CeIn$_3$/LaIn$_3$, the thickness reduction of the AFM heavy fermion CeIn$_3$ layers leads to suppression of the magnetic order and enhancement of the effective mass, both of which indicate an approach to a QCP. In the vicinity of the QCP, the resistivity exhibits a non-Fermi liquid behavior with a linear temperature dependence, which is consistent with the scattering from quantum fluctuations enhanced by the 2D-AFM QCP. Thus, we found that the dimensionality reduction of the Kondo lattice suppresses the 3D AFM state and surprisingly generates 2D AFM fluctuations instead of 3D ones. 

In the superconducting Kondo superlattices CeCoIn$_5$/YbCoIn$_5$, heavy fermion superconductivity is realized even at a single-unit-cell-thick layer of heavy fermion superconductor CeCoIn$_5$. The two-dimensionalization causes dramatic changes in the temperature and angular dependencies of the upper critical field. Moreover, we observed a substantial reduction of the magnetic fluctuations at the interface CeCoIn$_5$ layers neighboring the YbCoIn$_5$ layers using site-selective NMR measurements. These results can be well explained by the effect of the local ISB at the interfaces between CeCoIn$_5$ and YbCoIn$_5$ layers. Subsequent to this work, we have shown that the block-layer ISB, which results from the thickness modulation of the YbCoIn$_5$ layers, also affects the superconducting properties of the superlattices through the suppression of the Pauli effect. Thus, the entanglement of the local and block-layer ISBs, and the two-dimensionalization in the CeCoIn$_5$/YbCoIn$_5$ superlattices can help to achieve even more fascinating superconducting states. Further studies using various experimental techniques such as a combined STM-MBE system, site-selective NMR \cite{Yamanaka2015}, scanning SQUID microscopy, and terahertz spectroscopy \cite{Scheffler2013,Pracht2016} would be required to clarify how an exotic superconducting state is realized in this system.  

Finally, we provide an up-to-date status report concerning Kondo superlattices. Most recently, we successfully fabricated two new superlattices: tricolor superlattice CeCoIn$_5$/YbCoIn$_5$/YbRhIn$_5$, which does not possess global inversion symmetry, and a hybrid superlattice of heavy fermion superconductor CeCoIn$_5$ and AFM heavy fermion CeRhIn$_5$. At the time of writing this report, we have observed that the tricolor superlattices exhibit $H_{c2\perp}(T)$ and $H_{c2}(\theta)$ behaviors similar to those of local and block-layer ISBs, whereas the hybrid superlattices exhibit both $T_{\rm N}$ and $T_c$. These superlattices offer the potential of further in-depth exploration of the role of ISBs in  superconductivity and of the interplay between AFM fluctuations and superconductivity.

\begin{acknowledgments}
The authors acknowledge collaborations with H. Shishido, Y. Mizukami, T. Yamanaka, T. Watashige, K. Ishida, Y. Yanase, H. Kontani, T. Terashima, A. I. Buzdin, M. Yamashita, H. Ikeda, S. Yasumoto, D. Watanabe,  K. Yasu, T. Kato, R. Endo, and R. Kobayashi. We also thank the following for helpful discussions: M. Izaki, N. Kawakami, H. Kurata, K. Miyake, A. Tanaka, R. Arita, A. V. Chubukov, M. J. Graf, P. A. Lee, N. P. Ong, S. A. Kivelson, T. Takimoto, I. Vekhter, V. Mineev, R. Ikeda, S. Fujimoto, A. Balatsky, M. Sigrist, R. Peters, and Y. Tada. This work was supported by Grant-in-Aids for Scientific Research (S), and on Innovative Areas ``Topological Materials Science''.
\end{acknowledgments}

\bibliography{KondoSL_revised.bib}

\end{document}